\documentclass[runningheads]{llncs}
\usepackage{hyperref}
\usepackage[dvipsnames]{xcolor}
\usepackage[colorinlistoftodos,prependcaption,textsize=tiny]{todonotes}
\usepackage{graphicx}
\usepackage{microtype}
\usepackage[utf8]{inputenc}
\usepackage{algorithm}
\usepackage[noend]{algpseudocode}
\usepackage{tabularx}
\usepackage{footnote}
\makesavenoteenv{tabular}
\makesavenoteenv{table}
\usepackage{booktabs}
\usepackage{paralist}
\usepackage{etoolbox}
\usepackage{caption}
\usepackage{subcaption}
\usepackage{amssymb}
\usepackage{amsmath}
\usepackage{rotating}
\usepackage{booktabs}
\usepackage{multirow}
\usepackage{verbatim }
\usepackage{pifont}
\usepackage{enumitem}

\newcolumntype{P}[1]{>{\centering\arraybackslash}p{#1}}
\newcommand\footnoteref[1]{\protected@xdef\@thefnmark{\ref{#1}}\@footnotemark}

\newtheorem{prop}{Proposition}
\hypersetup{
	pdftoolbar=true,        
	pdfmenubar=true,        
	pdffitwindow=false,     
	pdfstartview={FitH},    
	pdftitle={Privacy-Preserving Directly-Follows Graphs: Balancing Risk and Utility in Process Mining},    
	pdfauthor={},     
	pdfsubject={},   
	pdfcreator={},   
	pdfproducer={}, 
	pdfkeywords={}, 
	pdfnewwindow=true,      
	colorlinks=true,       
	linkcolor=Brown,          
	citecolor=OliveGreen,        
	filecolor=magenta,      
	urlcolor=NavyBlue           
}

\makeatletter
\let\runauthor\@author
\let\runtitle\@title

\makeatletter
\providecommand{\subtitle}[1]{
	\apptocmd{\@title}{\par {\large #1 \par}}{}{}
}
\makeatother

\begin{document}
\title{Privacy-Preserving Directly-Follows Graphs: Balancing Risk and Utility in Process Mining}

\titlerunning{Privacy-Preserving Directly-Follows Graphs: Balancing Risk and Utility}
\author{Gamal Elkoumy\inst{1} \and
 Alisa Pankova\inst{2}\and
Marlon Dumas\inst{1}}
\authorrunning{G. Elkoumy et al.}

%
\institute{University of Tartu, Tartu, Estonia \\
	\email{\{gamal.elkoumy,marlon.dumas\}@ut.ee} \and
Cybernetica, Tartu, Estonia\\
\email{\{alisa.pankova\}@cyber.ee}
}

\sloppy

\maketitle              
\begin{abstract}


Process mining techniques enable organizations to analyze business process execution traces in order to identify opportunities for improving their operational performance. Oftentimes, such execution traces contain private information. For example, the execution traces of a healthcare process are likely to be privacy-sensitive. In such cases, organizations need to deploy Privacy-Enhancing Technologies (PETs) to strike a balance between the benefits they get from analyzing these data and the requirements imposed onto them by privacy regulations, particularly that of minimizing re-identification risks when data are disclosed to a process analyst.
Among many available PETs, differential privacy stands out for its ability to prevent predicate singling out attacks and its composable privacy guarantees. A drawback of differential privacy is the lack of interpretability of the main privacy parameter it relies upon, namely epsilon. This leads to the recurrent question of how much epsilon is enough?
This article proposes a method to determine the epsilon value to be used when disclosing the output of a process mining technique in terms of two business-relevant metrics, namely absolute percentage error metrics capturing the loss of accuracy (a.k.a.\ utility loss) resulting from adding noise to the disclosed data, and guessing advantage, which captures the increase in the probability that an adversary may guess information about an individual as a result of a disclosure. The article specifically studies the problem of protecting the disclosure of the so-called Directly-Follows Graph (DFGs), which is a process mining artifact produced by most process mining tools. The article reports on an empirical evaluation of the utility-risk trade-offs that the proposed approach achieves on a collection of 13 real-life event logs.

\end{abstract}

%
%
%

\section{Introduction}
\label{sec:introduction}

Process Mining is a family of techniques to analyze event logs generated by enterprise information systems in order to help organizations to identify opportunities to enhance the efficiency, compliance, and quality of their business processes~\cite{dumas2013business}. The primary input of a process mining technique is an event log, consisting of a collection of event records. Each record contains a reference to a process instance (the case identifier), a reference to an activity (activity label), and at least one timestamp (e.g.,\ the timestamp of the activity completion). Besides, each event may contain other attributes such as the resource (e.g.,\ worker) who performed the activity. An excerpt of an event log of a healthcare process is shown in Table~\ref{tbl:event_log}. In this event log, each case identifier refers to a patient, each activity label encodes a treatment, test, or another relevant event during a patient's hospital stay, and each resource refers to an employee (e.g., a doctor). For simplicity, timestamps are represented as integer numbers.

\begin{table}[hbtp]
\centering
\caption{Example of an event log of a healthcare process}
	\begin{tabular}[t]{lllll}
\hline

Case     &     Activity     &     Timestamp     &     Resource     &Additional Attributes\\\hline

P1    &    A    &    1    &    S1    &    .......  \\

P1    &    B    &    1.2    &    S2    &    .......  \\ 

P1    &    C    &    2.2    &    S3    &    .......  \\ 

P1    &    D    &    2.4    &    S4    &    .......  \\ 

P2    &    A    &    5    &    S1    &    .......  \\ 

P2    &    B    &    8    &    S2    &    .......  \\ 

P2    &    C    &    13    &    S3    &    .......  \\
P2    &    D    &    13.25    &    S4    &    .......  \\

P3    &    A    &    7    &    S1    &    .......  \\ 

P3    &    C    &    8    &    S3    &    .......  \\

P4    &    A    &    20    &    S1    &    .......  \\ 

P4    &    D    &    27    &    S4    &    .......  \\ 

P5    &    A    &    30    &    S1    &    .......  \\ 
...    &    ...    &    ...    &    ...    &    .......  \\ \hline
	\end{tabular}

\label{tbl:event_log}
\end{table}

Given an event log, process mining techniques can produce a variety of outputs. A typical output produced by process mining tools is the so-called \emph{Directly-follows Graph (DFG)} (a.k.a.\ \emph{process map}) of an event log. In addition to being a common approach to visualize the dependencies between activities in a process, the DFG is used as an intermediate artifact by various algorithms for automated discovery of process models~\cite{augusto2018automated}.  

\begin{figure*}[hbtp]
	\centering


	\begin{subfigure}[b]{0.7\textwidth}
		\centering
	\includegraphics[width=.98\columnwidth]{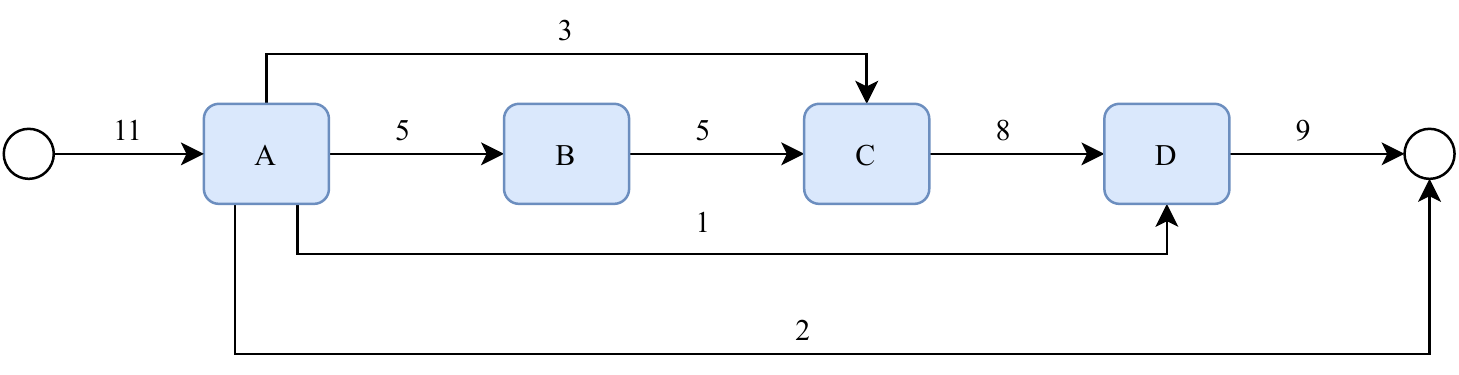}
	\caption{Frequency-annotated Directly-Follows Graph}
	\label{fig:dfg_graph}

	\end{subfigure}

~
	\begin{subfigure}[b]{0.7\textwidth}
	\centering
	\includegraphics[width=.98\columnwidth]{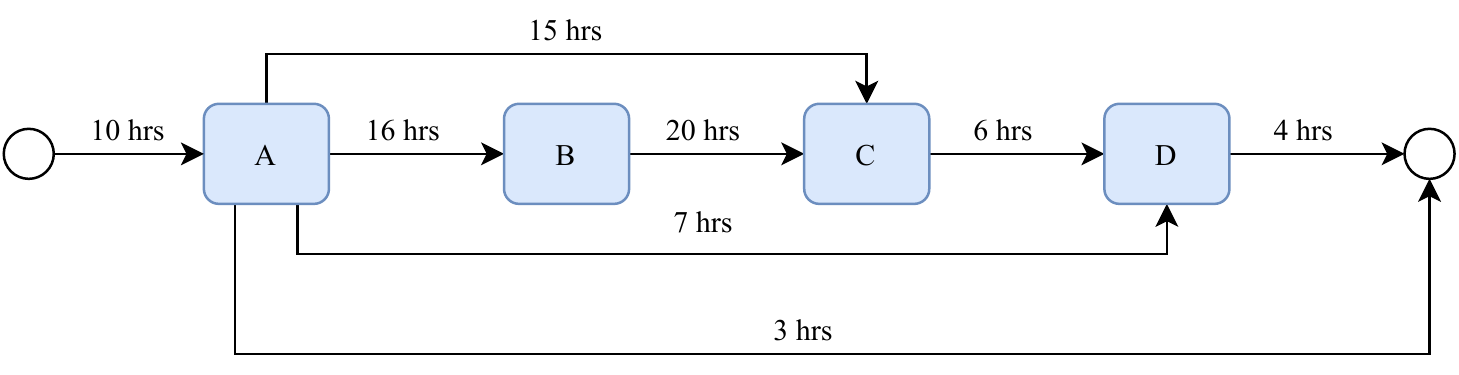}
	\caption{Timely-annotate Directly-Follows Graph with Maximum Aggregation}
	\label{fig:dfg_time}

\end{subfigure}
	\caption{ DFG of log in Table~\ref{tbl:event_log} }
	\label{fig:dfg}
\end{figure*}

A DFG is a directed graph in which each node represents activity in the process, and each arc represents a directly-follows relation between two activities, meaning that the activity that is the source of the arc was observed right before the activity that is the target of the arc at least once in the event log. Typically, each arc in the DFG is annotated with the number of times that the target activity immediately follows the source activity (arc frequency). However, it may also be annotated with other metrics such as the average timestamp difference between the source and the target activity or the maximum timestamp differences between the source and the target. For example, frequency-annotated DFG of the hospital event log mentioned above 
is shown in Figure~\ref{fig:dfg_graph}, and the timely-annotated DFG with maximum aggregation is shown in Figure~\ref{fig:dfg_time} . 




DFGs may contain sensitive information about individuals. For example, the DFG in Figure~\ref{fig:dfg_graph} contains 11 cases; only one case (P4) has a directly-follows relation (A, D). Let us consider the situation where an analyst happens to know that patient P4 underwent treatment D immediately after treatment A. Given this prior knowledge, the analyst is able to guess the time between the moment patient P4 arrived and left the hospital. 


Data minimization principles embedded in privacy regulations, such as  GDPR~\cite{GDPR} , 
 require that organizations put in place mechanisms to protect information about individuals when processing a dataset. 
The definition of individual data identification is articulated in Recital 26 of GDPR:

\textit{To determine whether a natural person is identifiable, account should be taken of all the means reasonably likely to be used, such as singling out, either by the controller or by another person to identify the natural person directly or indirectly.}




%




The notion of \emph{singling out} is elaborated in a guide by the Article 29 Data Protection Working Party~\cite{article29}. According to this guide, a person can be identified (singled out) using a dataset, if the dataset allows us to distinguish that person from all other persons represented in the dataset. In other words, a dataset allows a singling out of an individual (from a group) if there is a predicate that can be evaluated on that dataset and that uniquely distinguishes the  individual in question. The legal notion of singling out has been formalized by Cohen \& Nissim~\cite{cohen2020towards}. Specifically, Cohen \& Nissim give a mathematical formulation of a concept of Predicate Singling Out (PSO), capturing the idea that there exists a predicate that uniquely identifies a row in a dataset.


In the above example, the disclosure of the two DFGs in Figure~\ref{fig:dfg} allows for singling out. Specifically, the predicate ``undergoing treatment D immediately after treatment A'' allows us to single out an individual patient. Given these DFGs, an analyst can single out one of the patients if they happen to know that a given patient went from treatment A directly to treatment D. Given this prior knowledge, the analyst can then guess that this is the only patient who went from treatment A straight to D and, from the time-annotated DFG, they can guess the time this specific patient spent in the hospital (one hour).


To be able to analyze privacy-sensitive datasets within the framework of GDPR and similar privacy regulations, organizations need to make use of Privacy-Enhancing Technologies (PETs). Among many existing PETs~\cite{dwork2014algorithmic,wagner2018technical}, differential privacy stands out due  to the following properties:
\begin{itemize}
	\item It is PSO-secure as formalized in~\cite{cohen2020towards}. 
 \item It offers composable privacy guarantees~\cite{dwork2008differential,dwork2006calibrating}. Here, composability refers to the property wherein the total risk of a series of disclosures can be estimated from the risks of the separate disclosures. Specifically, the risk of disclosing the two DFGs in Figure~\ref{fig:dfg} is the sum of the risk of disclosing the frequency-annotated DFG and the risk of disclosing the time-annotated DFG. 
\end{itemize}
A differentially private disclosure mechanism operates by adding noise to the released data in order to achieve anonymization.
A drawback of differential-privacy is the lack of interpretability of the main privacy parameter it relies upon, namely $\epsilon$. This leads to the recurrent question of how much $\epsilon$ is enough?~\cite{lee2011much}


This article proposes a method to determine the $\epsilon$ value to be used when disclosing an annotated DFG in terms of two business-relevant metrics, namely: (i) \emph{absolute percentage error}, which captures the loss of accuracy resulting from adding noise to the disclosed data (a.k.a.\ utility loss); and (ii) \emph{guessing advantage}, which captures the increase in the probability that an adversary may guess information about an individual as a result of a disclosure. 

Specifically, the method proposed in this article addresses the following 
dual problems:
\begin{enumerate}[label=\textbf{P\arabic*}]
	\item\label{int:prob:1} Given an event log L and its corresponding DFG $G$, and given a maximum level of acceptable risk $\delta$, compute an optimal $\epsilon$ value such that the risk of disclosing the DFG $G'$ obtained by applying an $\epsilon$-differentially private mechanism to $G$, is below $\delta$. In this context, an optimal $\epsilon$ is an $\epsilon$ that minimizes the difference between DFGs $G$ and $G'$, measured via mean absolute percentage error.
	\item\label{int:prob:2} Given an event log L and its corresponding DFG $G$, and given a maximum level of mean absolute percentage error $\mu$, compute an optimal $\epsilon$ value such that the difference between $G$ and the DFG $G'$ obtained by applying an $\epsilon$-differentially private mechanism to $G$, is below $\mu$. Here, an optimal $\epsilon$ is an $\epsilon$ that minimizes the disclosure risk of $G'$.
\end{enumerate}
We tackle the above problem under the following requirements:
\begin{enumerate}[label=\textbf{R\arabic*}]
	\item\label{int:req:1}
	The noisified DFG $G'$ must have the same set of activities (nodes) as the original DFG $G$. The rationale for this requirement stems from the observations made by van der Aalst~\cite{van2019practitioner} that, when nodes are removed from a DFG, the semantics of the arcs in the resulting graph is no longer a ``directly-follows'' relation, and that this, in turn, may lead users to make incorrect conclusions about the performance or conformance of the process.
	\item\label{int:req:2}
	The risk metric is interpretable. In the absence of consensus on the interpretability of different privacy risk metrics~\cite{wagner2018technical}, we adopt a risk metric in the category of \emph{adversary's success probability}~\cite{laud2020framework,laud2019interpreting,wagner2018technical}, as this category of risk metrics can be directly translated into a probability of an adversary gaining knowledge about an individual's private information from one or more data disclosures.
\end{enumerate}

The article reports on an empirical evaluation of the computational efficiency of the proposed approach and the utility-risk trade-offs it offers on a collection of 13 real-life event logs.
	
The remainder of this article is structured as follows. In Section~\ref{sec:background}, we review related work. In Section~\ref{sec:approach}, we discuss the application of differential privacy to DFGs, and we present the attack model. In Section~\ref{sec:balance}, we propose a model to balance between the risk and the utility during DFG disclosure using differential privacy. In Section~\ref{sec:evaluation}, we discuss the empirical evaluation. Finally, Section~\ref{sec:conclusion} draws conclusions and outlines future work.


\section{Background and Related Work}
\label{sec:background}

In this section, we provide an introduction of the $\epsilon$-differential privacy mechanism and the global sensitivity of a privacy mechanism. We then give an overview of existing privacy models for process mining and event-log publishing.

\subsection{Differential Privacy}
A database $D$ is a set of data attributes whose values are drawn from a universe $U$. A tuple in a database belongs to an individual who needs his privacy to be protected. Each tuple is a collection of attributes $A= A_1, A_2, ..., A_m $, where $m$ is the number of attributes in the database.

The values of each attribute belong to an attribute domain. 
A mechanism $M : D \rightarrow \mathbb{R}^d$ is a function that maps a database $D$ to a vector of real numbers, distributed over a given range, herein denoted $Range(M)$. A privacy mechanism $M$ is an $\epsilon$-differentially private when adding or dropping a single data element in a database affects only the probability of the output by a small multiplicative factor \cite{lee2011much}.

\begin{definition} [$\epsilon$-differentially private mechanism]\label{def:dp1}
A mechanism $M$ is said to be $\epsilon$-differentially private if all the data sets $D_1$ and $D_2$ differing at most on one item, and on all $S \subseteq Range (M)$.

\centering $Pr[M(D_1) \in S] \leq exp(\epsilon) \times Pr[M(D_2) \in S]$
\end{definition}

In this article, we consider an interactive privacy mechanism as in \cite{lee2011much}, where a user issues a query to a database and receives a noisified response. Given a query function $f$, the amount of noise to be added depends on the \emph{sensitivity} of $f$. The sensitivity of query function $f$ is the largest difference in the output of $f$ that can occur when we vary the input of $f$ by one single data element.

\begin{definition} [Global Sensitivity] \label{def:gs1}
	For a query function $f : D \rightarrow \mathbb{R}^d$, the global sensitivity is
	
	\begin{center}$\Delta f= \max\limits_{D_1,D_2} |f(D_1) - f(D_2)|$ \end{center}

  $\forall D_1,D_2 $ differing in one item at most.
\end{definition}

A common practice in the field of differential privacy is to add noise drawn randomly from the Laplace distribution. The Laplace distribution $Lap(\lambda, \mu)$ has a density function $h(x) = \frac{1}{2\lambda} exp(-\frac{|x-\mu|}{\lambda} )$ where $\lambda$ is a scale factor and $\mu=0$ is the mean. For a given database $D_1$ and a query function $f$, a randomized mechanism $M_f$ gives $\epsilon$-differential privacy if it returns $f(D_1)+Y$ as an outcome where $Y$ is chosen i.i.d from   $Lap(\frac{\Delta f}{\epsilon},0)$ \cite{dwork2006calibrating,dwork2008differential,dwork2014algorithmic}.

Global sensitivity may turn out to be very high even if $D_1$ and $D_2$ differ only in one item, taking into consideration all the attributes of that item. For example, if we are computing SUM over an attribute $A$, then adding or removing an item affects the sum as much as the maximum possible value of this attribute. If we are interested in hiding $A$, we may require differential privacy for $D_1$ and $D_2$ concerning attribute $A$, i.e. $D_1$ and $D_2$ have the same number of tuples but are different in the attribute $A$ of one item.

\begin{definition} [$\epsilon$-differentially private mechanism w.r.t. attribute]\label{def:dp2}
A mechanism $M$ is said to be $\epsilon$-differentially private w.r.t. attribute $A$ if all the data sets $D_1$ and $D_2$ differing at most on one item and on all $S \subseteq Range (M)$.
\centering $Pr[M(D_1) \in S] \leq exp(\epsilon \cdot |D_1.A - D_2.A|) \times Pr[M(D_2) \in S]$
\end{definition}

We can use the Laplace mechanism to achieve DP w.r.t. attribute using global sensitivity w.r.t. attribute.

\begin{definition} [Global Sensitivity w.r.t. attribute] \label{def:gs2}
For a query function $f : D \rightarrow \mathbb{R}^d$, the global sensitivity w.r.t. attribute $A$ is
	
	\begin{center}$\Delta f= \max\limits_{D_1,D_2} \frac{|f(D_1) - f(D_2)|}{|D_1.A - D_2.A|}$ \end{center}

  $\forall D_1,D_2 $ differing in one item at most.
\end{definition}

The $\epsilon$-differential privacy limits the ability of an adversary to identify an individual (using Definition~\ref{def:dp1}) or determining the value of a particular attribute (using Definition~\ref{def:dp2}). Although, for small values of $\epsilon$ , the utility of the outcome is reduced. 
Lee et al. \cite{lee2011much} study the interpretation of the differential privacy parameters. They consider the re-identification probability of individuals in a database. They demonstrate the challenges with choosing $\epsilon$ to protect individual information with a fixed probability. Although the parameter $\epsilon$ is used to quantify the risk of releasing a statistical analysis of sensitive data, it is not an absolute metric of privacy, but rather, a relative value. Hsu et al. \cite{hsu2014differential} present an economical method for choosing $\epsilon$. They assume that there are two conflicting goals: learning the correct analysis from the data and keeping the data of individuals private. They use a privacy budget for individuals to represent the maximum loss of privacy that they are willing to accept. Based on the privacy budget, they derived formulas on privacy parameters to keep the balance between the conflicting objectives. Laud et al. \cite{laud2020framework,laud2019interpreting} provide a method of choosing the value of $\epsilon$ based on attacker's advantage in achieving a particular goal. In their model, they state the attacker's goal as a Boolean expression of guessing attributes. They studied the change of prior and posterior probabilities of guessing. This method is related to the privacy metrics: adversary's success probability~\cite{wagner2018technical,rastogi2007boundary,nergiz2007hiding}, and accuracy of the adversary's estimate~\cite{yang2013towards,kantarcioglu2004data,cheng2006preserving}.  In this article, we adopt the guessing advantage mechanism to build differential private DFGs extracted from event logs.


\subsection{Privacy-Preserving Process Mining}

Previous studies on privacy-preserving process mining have addressed two complementary concerns. The first concern is how to ensure confidentiality and privacy during the computation of process mining operations, particularly in a cross-organizational setting. For example, Tillem et al.~\cite{tillem2016privacy,tillem2017mining} propose a distributed secure processing protocol for discovering process models using the alpha algorithm. Similarly, Elkoumy et al.~\cite{elkoumy2020secure,elkoumy2020shareprom} propose a method for securely computing the DFG of an event log in an inter-organizational setting. The second concern is how to ensure that the disclosure of an event log or of a derivative thereof (e.g.\ a DFG) complies with privacy requirements. This article, and the rest of our review of related work, focuses on this latter concern.

Rafiei et al.~\cite{rafiei2018ensuring} define a pseudonymization technique based on masking, i.e.\ replacement of attribute values with pseudonyms. Masking has the advantage that it does not affect the utility (accuracy) of aggregate queries, including those queries required to compute the frequency-annotated or time-annotated DFG. On the other hand, it is well-known that masking does not provide privacy risk guarantees~\cite{fahrenkrog2019pretsa}. As such, Rafiei et al.~\cite{rafiei2018ensuring} do not address the problem of controlling a disclosure in order to ensure a certain level of disclosure risk and so, they neither address problem ~\ref{int:prob:1} nor ~\ref{int:prob:2} as posed in Section~\ref{sec:introduction}. 


PRETSA~\cite{fahrenkrog2019pretsa} proposes a technique to ensure the k-anonymity of event logs through suppression (e.g., removal of events or cases). This technique addresses a similar problem to~\ref{int:prob:1}, but it uses k-anonymity instead of differential privacy.  K-anonymity is an interpretable risk metric (as ``k'' corresponds to the minimum size of a group in the dataset). As such, this approach fulfills requirement~\ref{int:req:2} spelled out in Section~\ref{sec:introduction}. However, k-anonymity is neither composable nor secure against singling out attacks. Indeed, Cohen et al. \cite{cohen2020towards} proved that under some settings, k-anonymity enables an attacker to perform a predicate single out attack with a probability of 37\%. The fact that PRETSA uses suppression implies that the DFG computed from an anonymized log might contain fewer nodes than the original DFG, and hence this technique does not fulfill requirement~\ref{int:req:1}.

Rafiei et al.~\cite{rafiei2020tlkc} propose another approach for publishing event logs using a k-anonymity mechanism. They use a privacy model called TLKC, where risk is measured via a confidence measure and is dependent on background knowledge (specifically the length of a sequence). Their approach covers both frequency and time queries. TLKC uses the \emph{minimum trace-frequency} as a measure of utility. It tries to maximize the frequent traces in the anonymized event log. Like PRETSA,  this approach suppresses entire traces from the log, and hence, the DFG of the anonymized log may contain fewer nodes (activities) than the DFG of the original log. In some settings, the percentage of deleted nodes is high. For example, when applying this technique to an event log of a patient treatment process  (specifically the Sepsis log~\cite{felix2017sepsis}), under a strong privacy setting, the DFG of the anonymized log contains only 13\% of the activities in the original event log. As such, this approach does not fulfill requirement~\ref{int:req:1}. On the other hand, it relies on interpretable notions of risk. Therefore, it fulfills requirement~\ref{int:req:2}. This approach does not address problem~\ref{int:prob:2} -- it does not seek to optimize disclosure risk given a desired level of utility. The reliance on k-anonymity implies that this approach does not provide composable privacy guarantees nor does it protect against PSO attacks.


Mannhardt et al.~\cite{mannhardt2019privacy} propose a differential privacy framework for two types of queries over event logs: the query that computes the frequency-annotated DFG from an event log, and the query that computes the distribution of distinct execution traces (case variants) from an event log. The authors empirically study the impact of varying the $\epsilon$ parameter on the accuracy (recall and precision) on multiple event logs. However, they do not propose an approach to optimize the utility (accuracy) given a desired risk level ($\epsilon$) nor the other way around and hence this approach does not address problem~\ref{int:prob:1} nor~\ref{int:prob:2}.


 
 

PRIPEL~\cite{fahrenkrog2020pripel} is another approach for protecting the disclosure of DFGs using differential privacy mechanism. While Mannhardt et al.~\cite{mannhardt2019privacy} focus on frequency-annotated DFGs, PRIPEL additionally deals with time-annotated DFGs. However, PRIPEL does not seek to optimize the $\epsilon$ given a utility level nor vice-versa and thus it neither addresses problem~\ref{int:prob:1} nor~\ref{int:prob:2}.




Other studies on privacy-preserving process mining fall outside the scope of the present article, as they do not provide mechanisms for protecting the disclosure of an event log or a DFG. For example, Pika et al. \cite{pika2019towards} analyze privacy requirements for process mining in the healthcare domain. They note the inherent trade-off between privacy and utility of the outputs in this context and advocate the use of differential privacy to balance this trade-off. However, they do not propose a concrete disclosure control mechanism.  

Rafiei et al.~\cite{rafiei2018ppdp} extend the XES standard and provide a formal definition of the operation used by anonymization privacy models in privacy-preserving process mining. Again, they do not define any specific disclosure control mechanism. Other studies deal with the quantification of disclosure risk. In this line, von Voigt et al.~\cite{von2020quantifying} propose a method to quantify the re-identification risk entailed by an event log disclosure using a measure of individual uniqueness. Meanwhile, Rafiei et al.~\cite{rafiei2020towards} introduce quantification methods for both the disclosure risk and data utility, but without proposing a concrete disclosure control mechanism.

To summarize, only two previous studies address problem~\ref{int:prob:1} are PRETSA~\cite{fahrenkrog2019pretsa} and TLKC~\cite{rafiei2020tlkc}. However, these approaches do not provide composable privacy guarantees, nor do they protect against PSO attacks due to their reliance on k-anonymity. They do not address problem~\ref{int:prob:2} (relating utility back to risk) and they do not fulfill requirement~\ref{int:req:1} (they may suppress nodes from the DFG).

\newcommand{\share}[1]{[\![{#1}]\!]}

\section{Preliminaries}
\label{sec:approach}
In this section, we define the privacy quantification for both risk and utility to address the problems mentioned in Section~\ref{sec:introduction}. First, we start with the formal definition of a differentially-private DFG. Then, we present the utility measure used in this article as the absolute percentage error. Moreover, 
we present our attack model, and we define the risk measure as the probability of guessing advantage in order to provide an interpretable risk quantification to fulfill Requirement~\ref{int:req:2}.  

\subsection{Differentially-Private DFGs}

The events that happen inside an organization are being stored inside an event log. The basic event log should contain an identifier for each case, the activity executed, and the timestamp at which the event happened. The event log can contain extra attributes such as the resource executed the activity, the department supervised the activity execution, etc.

\begin{definition} [Events, Traces, Event Logs] 
	An event log $L= \{e_1, e_2, ..., e_n\}$ is a set of events $e=(i,a,ts,c_1,....,c_k)$, each capturing an execution of an activity $a\in A$, at timestamp $ts \in T$, as part of a case $i \in I$, and it contains a list of additional attributes $c_1, c_2, ....,c_k$ where $c_j \in C_j$ and $1 \leq j \leq k$. A trace $t=\langle e_1, e_2, ..., e_m\rangle$ is a set of events that are ordered by  their timestamps to represent a single execution of the process. All events from the same trace have the same identifier $i$.
\end{definition}

Given an event log, we seek to protect the disclosure of a DFG of this log, as defined below.






\begin{definition}[Directly-Follows Graph]
	The directly-follows graph of an event log L is	 $G(L)= \{ (D_1,R_1), (D_2,R_2), .....,(D_{n},R_{n}) \}$, with:
	\begin{itemize}
	 \item $D=\{ (x,y) | x \in A  \wedge  y \in A  \wedge (x <_L y) \}$ is a pair of two activities and,
	 \item $R$ is the set of edges of the graph with weights $W=\{w|w \in \mathbb{R}\}$  and,
	\item $x <_L y$ iff $\exists t=\langle e_1,e_2, ...,e_m \rangle $and  $ i \in \{ 1,..., m-1\}$ such that $t\in Traces(L)$ and $ t_i.a =x$ and $t_{i+1}.a = y$ 
	
	\end{itemize}
	
\end{definition}

The weights $W$ on the edges of the DFG are the result of an aggregate query function $f$, that represents the directly-follows relation between the pair of values $D$. An aggregation function $f$ can be the count of occurrences as in the frequency annotated DFG, or an aggregation function over the time differences between the pair of values, e.g., the max time difference 
, as in the time annotated DFG. For example, consider the event log in Table~\ref{tbl:event_log}, the corresponding DFG is given in Figure~\ref{fig:dfg_graph}. This DFG shows the frequency as the weights in the DFG.

We seek to protect DFGs by adding noise to the weights on the edges of the DFG in such a way that we can ensure a certain level of differential privacy. The output of our technique is thus a differentially-private DFG as defined below.

\begin{definition}[Differentially-Private DFG]
	Let $M$ be an $\epsilon$-differentially private mechanism that uses Laplacian distribution. Let $M_f$ be a query function that computes the directly follows relation between a pair of two values $x,y$ over the set of activities $A \in L$ and adds Laplace noise to result.  A differentially-private directly-follows graph is defined as
	
	$G_P = M(G)= \{(D_1,M_f(R_1)),  (D_2,M_f(R_2)), ....., (D_{n}, M_f(R_{n}))\}$ 
	\end{definition}

To apply differential privacy to a DFG, we view the DFG as a histogram with one value per edge in the DFG. For example, the DFG in Figure~\ref{fig:dfg_graph} is shown as a histogram in Table~\ref{tbl:seq}. We write ``--'' to refer to the start of a case and the end of a case. A differentially-private DFG is computed by applying a mechanism that adds a noise from a Laplacian distribution. To add noise, we need to alter some elements in the DFG. In this article, we focus on altering the weights of the DFG, in such a way that the altered values are still strictly positive numbers. We do not consider other possible alterations such as adding or deleting edges. 

\begin{table}[hbtp]
	\centering	
	\caption{DFG of log in Table~\ref{tbl:event_log} represented as a histogram of frequencies/time differences}
	\begin{tabular}[t] { p{2cm}	p{2cm}	p{2cm}} 
		\hline
		
		Relation     &     Frequency     &     Sum of Time Differences     \\\hline 
		(--,A)    &    11    &    34   \\ 

		(A,B)    &    5    &    39.2      \\  

		(A,C)    &    3    &    22       \\  

		(A,D)    &    1    &    7     \\  

		(B,C)    &    5    &    52      \\  

		(C,D)    &    8    &    19.3     \\ 


		(D,--)    &    9    &    12.2        \\  

		(A,--)    &    2    &    3.5         \\ \hline 
		
	\end{tabular}
	
	\label{tbl:seq}
\end{table}

\subsection{Accuracy Measure}

To measure the impact of noise injection on a DFG, we consider the Absolute Percentage Error (APE) for every edge and Mean Absolute Percentage Error (MAPE) and Symmetric Mean Absolute Percentage Error (SMAPE).

\begin{definition}[APE, MAPE, and SMAPE] \label{def:mape}

For a DFG with values on its edges, the Absolute Percentage Error (APE) for every edge, and Mean Absolute Percentage Error (MAPE) and Symmetric Mean Absolute Percentage Error (SMAPE) are 

\centering
$APE=\left|\frac{A - F}{A}\right|$ \label{df:MAPE}

$MAPE=\frac {1}{m} \sum_{i=1}^{m} \left|\frac{A_i - F_i}{A_i}\right|$

$SMAPE=\frac {1}{m} \sum_{i=1}^{m} \left|\frac{A_i - F_i}{A_i + F_i}\right|$

where $A_i$ is the actual value, $F_i$ is the noisified value and $m$ is the number of edges.
	
\end{definition}

An Absolute Percentage Error (APE) is 
the difference between the actual value $A$ and the noisified value $F$ as a percentage of the actual value of an edge of the graph. The Mean Absolute Percentage Error (MAPE) measures the difference across the entire graph as the average value of APE. MAPE gives a general measurement of the effect of noise on the entire graph. Symmetric Mean Absolute Percentage Error (SMAPE) is the absolute difference between $A_i$ and $F_i$ divided by the absolute sum of values $A_i$ and $F_i$. Its values then are summed and divided by the number of edges. We include the SMAPE in our measurements as it is not susceptible to large outliers in real-world event logs \cite{tofallis2015better}.


	
	
	

\subsection{Privacy Threats and Risk Measure}


In this article, we consider a scenario of data collection and DFG publishing. We assume that a process model holder (e.g., hospital) intends to disclose the model to the data analyst who is not trustworthy. The data analyst may perform actions to identify sensitive information from DFG. 

	

Specifically, we consider the case where an attacker (in our setting, the analyst) who has access to a DFG would try to use this DFG to guess an individual's private information. In particular, we consider the following attacker goals:
\begin{enumerate}
\item\label{att:goal:1} Has the individual been incorporated in a particular directly-follows relation? The output is a single bit that has an exact value $\in \{0,1\}$.
\item\label{att:goal:2} How much time did it take for a particular directly-follows relation? The output is a non-negative real value $\in \mathbb{R}_0^{+}$ that is allowed to be guessed with absolute precision.
\end{enumerate}


To mitigate these attacks, we use a differential privacy mechanism. However, 
differential privacy relies on a parameter ($\epsilon$) which does not have a straightforward  interpretation, we provide a layer on top of differential privacy that allows a data owner to control the trade-off between the utility of the data disclosed and the associated disclosure risk. This disclosure risk is captured as a notion of the attacker's advantage.

To this end, we rely on the work of Laud et al. \cite{laud2020framework,laud2019interpreting}, who proposed a framework that quantifies the attacker’s advantage as the difference between the prior and the posterior beliefs on the attribute he is trying to guess. In their framework, they work with general bounds and a generalized distribution of noise. In this article, we consider only the Laplace noise distribution. Let $g$ be the set of directly-follows relations before aggregation. The attacker has a goal $h(g)$, that models the information he is interested in guessing about $g$. In our case, the attacker has two goals: $h_1(g) \in \{0,1\}$ returning whether the targeted individual has been incorporated in a particular directly-follows relation of $g$, and $h_2(g) \in \mathbb{R}_0^{+}$ returning the time that the individual has spent in a particular relation of $g$. 

An attacker may have a prior knowledge about a group of individuals in the event log, e.g., a group of patients has taken a specific duration at a hospital. Let $k(g)$ represent the knowledge an attacker already has about $g$. Even without observing the DFG, the attacker can use his prior knowledge $k(g)$ to reveal information about an individual. The attacker's guess is considered successful if it falls into a subset $H_p$ of possible valuations of $h(g)$ that are considered ``good enough'', i.e. the value is guessed with a certain precision $p$. In the following, let $G$ be the random variable that corresponds to the distribution of $g$.





\begin{definition}[Prior Guessing Probability]\label{def:prior}
An attacker with a prior knowledge $k(g)$ 
has a successful guessing probability

	\centering \begin{equation}\label{eq:defp}
		P := Pr[h(G) \in H_p\ |\ k(G)=k(g)]\enspace.
	\end{equation}

\end{definition}

In our case, we assume a strong attacker who already knows everything about all other individuals covered by DFG, so $k(g)$ contains all the other directly-follows relations and their time differences except for the individual that the attacker is guessing. The attacker wants to guess information about individual with a certain precision $p$. 

Therefore, in our approach, the data publisher has to specify the precision $p$ at which he wishes to protect the time differences between consecutive events in the log. This precision could be expressed as an absolute value (e.g. 5 time units) or as a percentage of the total range of time durations. We adopt the latter approach. In other words, the precision is captured as a percentage of the total range of time durations observed in the log. Let us assume that the range of time differences is in the range between 0 and 100 time units. A precision of 0\% means that the data publisher wishes to protect the time differences within a range of $\pm 0$, which means that the data publisher only wishes to prevent the attacker from guessing the exact value of the time difference between every pair of consecutive events. Conversely, a precision of 1 means that the data publisher wishes to protect the time difference within a range of $\pm 100$. Similarly, a precision of 0.5 means that the data publisher wishes to prevent the attacker from guessing the time difference of any pair of events within a range of $\pm 50$ from the actual value. When the precision is zero, the attacker's guessing task is challenging (guessing the precise value). When the precision is 1 the attacker's guessing task is trivial (any guess would be within the desirable $\pm 100$ time units range).

For each occurrence ($e_1$, $e_2$) of a directly-follows relation (A, B) in the log, the data publisher intends to protect the time difference between event $e_1$ and event $e_2$. For each such occurrence , we assume that we have to protect this time difference between $e_1$ and $e_2$ from the strongest possible attacker, which is the attacker who knows the time differences of all other occurrences of the directly-follows relation (A, B) except for ($e_1$, $e_2$) and whose objective is to guess the time difference of ($e_1$, $e_2$) within the critical level of precision $p$. Note that this corresponds to a predicate singling-out PSO attack discussed in Section~\ref{sec:introduction}.

Given a level of guessing precision set by the data publisher, we will demonstrate in the next section that it is possible to calculate the prior probability that an attacker can guess the value of the time difference between any pair of consecutive events $e_1$ and $e_2$ when faced to the strongest possible attackers described above. 
 This prior probability is calculated by inspecting each possible value of the time difference TD between consecutive pairs of events, and (based on the distribution of TDs), determining the probability that the strongest attacker described above is able to guess TD with the level of precision set by the user.

The guessing advantage is the additional successful guessing probability that an attacker would gain after publishing an anonymized DFG, $M_f(g)$. In other words, the guessing advantage is the difference between the two probabilities: posterior probability (after observing $M_f(g)$ ) and prior probability (before observing the $M_f(g)$) of an attacker returning a value in $H_p$, with the assumption that the attacker’s knowledge $k(g)$ is a condition for both probabilities. Let $\delta$ be the maximum allowance for the guessing advantage. 

 


\begin{definition}[Guessing Advantage]\label{def:ga}
Attacker's advantage in achieving the goal $h$ with precision $p$ w.r.t. prior knowledge $k$ is at most $\delta$ if 
\begin{equation*}\label{eq:ga}
Pr[h(G) \in H_p\ |\ M_f(G)=M_f(g), K] - Pr[h(G) \in H_p\ |\ K] \leq \delta
\end{equation*}
where $K := (k(G)=k(g))$.
\end{definition}




The publisher can decide the parameter $\delta$, representing the upper bound of the guessing probability that the attacker will gain after publishing the DFG.


As mentioned above, the attacker has two goals. For attacker goal~\ref{att:goal:1}, we define $H_p := \{h_1(g)\}$ for any $p$, i.e the attacker needs to guess precisely whether the individual has been incorporated in a particular directly-follows relation. For attacker goal~\ref{att:goal:2}, we define $H_p := \{h' \in \mathbb{R}_0^{+}\ :\ |h_2(g) - h'| \leq p \cdot r\}$, where $r$ is the maximum value of the range of input values, i.e. a guess $h'$ is considered correct if its distance from the true value $h_2(g)$ is at most $p \cdot r$.

 

From \cite{laud2020framework,laud2019interpreting}, we get a formula for computing $\epsilon$ such that $\epsilon$-DP mechanism $M_f$ satisfies Definition~\ref{def:ga}. For attacker goal~\ref{att:goal:1}, the mechanism should be $\epsilon$-DP according to Definition~\ref{def:dp1}, and for attacker goal~\ref{att:goal:2}, it should be $\epsilon$-DP w.r.t. time attribute according to Definition~\ref{def:dp2}. Laud et al.~\cite{laud2020framework,laud2019interpreting} proposed an estimation of the posterior guessing probability. In this article, we apply the results of Laud et al.~\cite{laud2020framework,laud2019interpreting} to the disclosure of DFGs.

\begin{prop}[Posterior Guessing Probability~\cite{laud2020framework,laud2019interpreting}]
	The posterior guessing probability of an attribute ranging between 0 and r for a single individual
	 after the disclosure of a DFG is bounded by
	
	$P' \leq \frac{1}{1 + exp(-\epsilon \cdot r)\frac{1-P}{P}}\enspace.$
	
\end{prop}

\begin{proof}
			(Taken from~\cite{laud2020framework,laud2019interpreting}) Using the equality $Pr[X=x] = \sum_{y \in Y} Pr[X=x,Y=y]$ and Bayesian formula $Pr[A,B] = Pr[A|B]\cdot Pr[B]$, we can rewrite
	\begin{eqnarray*}\label{eq:prior}		
		P' &:=& Pr[h(G) \in H_p\ |\ M_f(G)=M_f(g), k(G) = k(g)]\\
		& = & \frac{Pr[h(G) \in H_p, M_f(G)=M_f(g), k(G) = k(g)]}{Pr[M_f(G)=M_f(g), k(G) = k(g)]}\\
		& = & \frac{\sum_{g': h(g') \in H_p, k(g) = k(g')}\ Pr[M_f(g')=M_f(g)]\cdot Pr[G = g']}{\sum_{g': k(g) = k(g')} Pr[M_f(g')=M_f(g)]\cdot Pr[G = g']}\\
		& = & \frac{1}{1 + \frac{\sum_{g':\ h(g') \notin h(g), k(g') = k(g)} Pr[M_f(g')=M_f(g)]\cdot Pr[G = g']}{\sum_{g'':\ h(g'') \in h(g), k(g'') = k(g)} Pr[M_f(g'')=M_f(g)]\cdot Pr[G = g'']}}\enspace,	
	\end{eqnarray*}
	For an $\epsilon$-DP mechanism $M_f$, since $g'$ and $g''$ differ in one item due to the condition $k(g')=k(g)=k(g'')$, we have $\frac{Pr[M_f(g')=M_f(g)]}{Pr[M_f(g'')=M_f(g)]} \geq exp(-\epsilon\cdot r)$, where $r$ is the largest possible difference between two values of an attribute that the attacker is guessing. This gives us	
	\begin{eqnarray*}
		P' &\leq& \frac{1}{1 + exp(-\epsilon\cdot r)\frac{\sum_{g':\ h(g') \notin h(g), k(g') = k(g)} Pr[G = g']}{\sum_{g'':\ h(g'') \in h(g), k(g'') = k(g)} Pr[G = g']}}\\
		&=& \frac{1}{1 + exp(-\epsilon\cdot r)\frac{Pr[h(G) \notin H_p, k(G) = k(g)]}{Pr[h(G) \in H_p, k(G) = k(g)]}}\\
		&=& \frac{1}{1 + exp(-\epsilon\cdot r)\frac{Pr[h(G) \notin H_p\ |\ k(G) = k(g)]}{Pr[h(G) \in H_p\ |\ k(G) = k(g)]}}\\
	\end{eqnarray*}
Putting $P$ as in Definition~\ref{def:prior}, we get
\begin{equation}\label{eq:postpr}
	P' \leq \frac{1}{1 + exp(-\epsilon \cdot r)\frac{1-P}{P}}\enspace.
\end{equation}

\end{proof}

Given Definition~\ref{def:ga}, $\delta$ is the maximum guessing advantage probability after disclosing the DFG. Therefore,
\begin{equation}\label{eq:guessing_adv_upper}
	P' - P \leq \delta \enspace.
\end{equation}
Substituting Eq~(\ref{eq:postpr}) into Eq~(\ref{eq:guessing_adv_upper}), we can estimate the the largest possible $\epsilon$ that achieves the upper bound $\delta$ as
\begin{equation}\label{eq:maineq}
\epsilon = \frac{-\ln\left(\frac{P}{1-P} \cdot \left(\frac{1}{\delta+P} - 1 \right) \right)}{r}\enspace.
\end{equation}
If we do not know the distribution of inputs and cannot compute $P$ precisely, we can take the ``worst-case'' probability $P'$ that maximizes the noise required to achieve guessing advantage $\delta$. This value has been computed in~\cite{laud2019interpreting} by minimizing $\epsilon$ in Eq(~\ref{eq:maineq}) w.r.t. $P$, which gives us
\begin{equation}\label{eq:worstcasep}
P' := \mathsf{argmin}_{P \in [0,1]}\frac{-\ln\left(\frac{P}{1-P} \cdot \left(\frac{1}{\delta+P} - 1 \right) \right)}{r} = \frac{1-\delta}{2}\enspace.
\end{equation}






%

\section{Balancing Risk and Utility}
\label{sec:balance}

In this section, we present our mathematical model to achieve a balance between the risk and utility. We present our analysis to address the problems mentioned in Section~\ref{sec:introduction}. To address both~\ref{int:prob:1} and~\ref{int:prob:2}, we relate the $\epsilon$-differential privacy parameter to both the risk and utility quantification.
We assume that the process map publisher has a maximum acceptable risk (guessing advantage) level that needs to be maintained, and we estimate an optimal differential-privacy parameter, $\epsilon$, and the percentage error to address problem~\ref{int:prob:1}. For Problem~\ref{int:prob:2}, we assume that the publisher has a minimum utility loss level (percentage error) that he needs to maintain, within a maximum acceptable risk (guessing advantage) level. We provide the mathematical model that enables the publisher to optimize the amount of injected noise through optimizing $\epsilon$, to achieve the balance between the desired utility and the acceptable risk.

\subsection{P1: From Risk to $\epsilon$ and MAPE}
Following, we present the mathematical model of Problem~\ref{int:prob:1}, mapping an input risk measure (guessing advantage) to both the differential privacy, $\epsilon$, and the mean average percentage error, MAPE. First, we start with the time annotated DFG, and then we discuss the frequency annotated DFG.
\paragraph{Time Annotated DFG.}

\begin{figure*}
	\begin{subfigure}[b]{0.45\textwidth}
		\centering
		\includegraphics[width=.98\columnwidth]{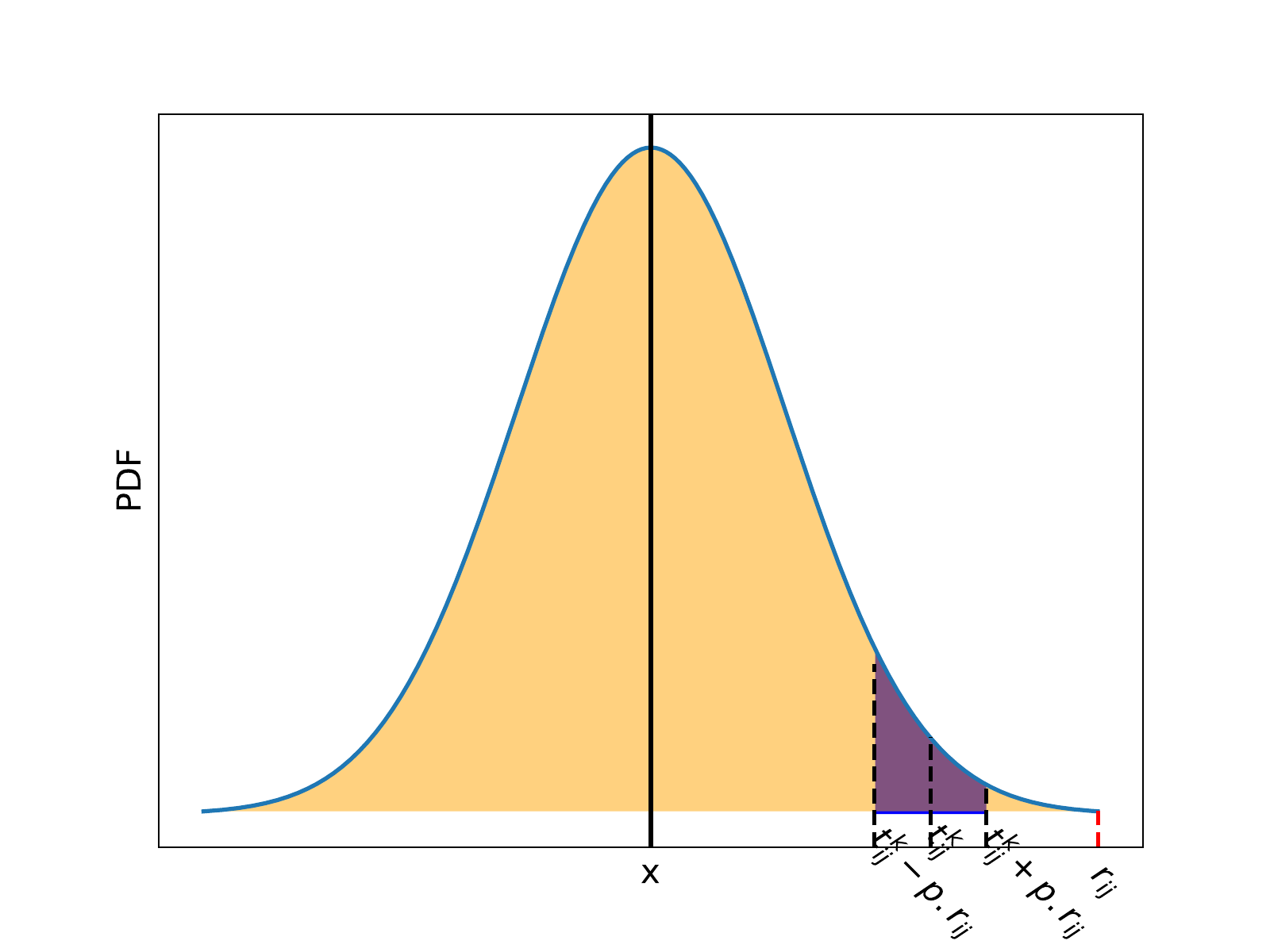}
		\caption{Prior probability of guessing a value $t^k_{ij}$ with precision $p$}
		\label{fig:guessing_advantage}
		
	\end{subfigure}
	\begin{subfigure}[b]{0.45\textwidth}
		\centering
		\includegraphics[width=.98\columnwidth]{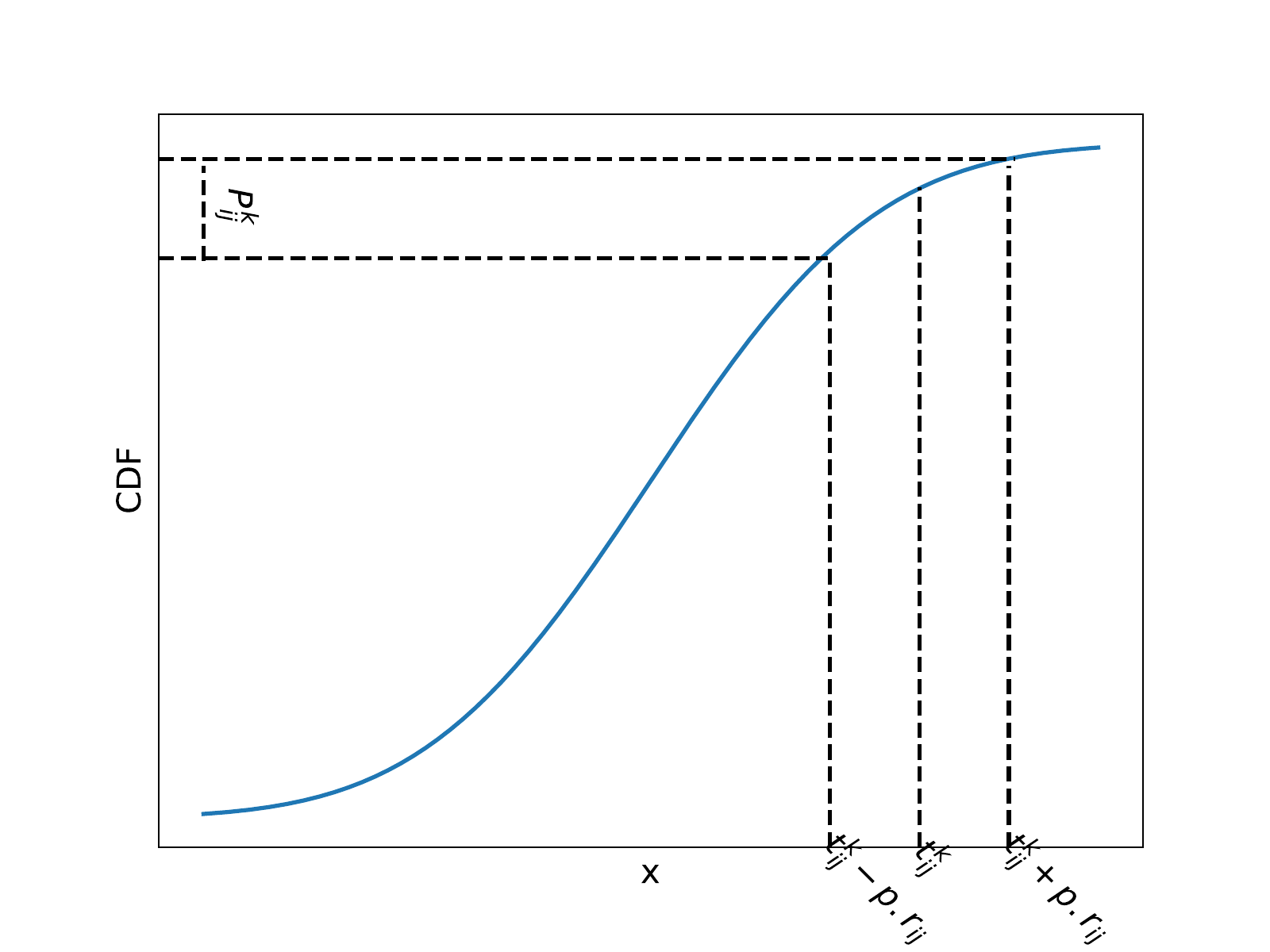}
		\caption{Calculating Prior Probability using CDF }
		\label{fig:CDF}
		
	\end{subfigure}
	\caption{ Calculating the Prior Probability $P^k_{ij}$ }
	\label{fig:guessing}
\end{figure*}

Every directly-follows relation of the time annotated DFG has more than one occurrence. 
Hence, for every directly-follows relation there is a distribution of values, and we apply an aggregation function, e.g. sum and maximum, over the values to determine the annotation of directly-follows relation. This distribution can be different for each directly-follows relation. Therefore, for each edge $(i,j)$, we estimate a prior guessing probability $P^k_{ij}$, as defined in Section~\ref{sec:approach}, for each occurrence $k$, where $i$ is the input activity, and $j$ is the output activity of the edge, as shown in Figure~\ref{fig:guessing_advantage}. We estimate $P^k_{ij}$ using Eq.~(\ref{eq:defp}).
In particular, we estimate the prior probability $P^k_{ij}$, for every occurrence with a true value $t^k_{ij}$, as the probability of guessing the value within the range of values $t^k_{ij} \pm p \cdot r_{ij}$, 
where $r_{ij}$ is the maximum possible value of time for the edge $(i,j)$, and $p$ is the precision. 
 To estimate the prior probability, we use the cumulative density function (CDF), as shown in Figure~\ref{fig:CDF}.

 Hence, for any distribution of the values having a cumulative density function (CDF), the prior guessing probability can be computed as
\begin{equation}\label{eqn:p_k}
P^k_{ij} = CDF(t^k_{ij} + p \cdot r_{ij}) - CDF(t^k_{ij} - p \cdot r_{ij}).
\end{equation}
In the cases when CDF cannot be estimated (e.g. there is too little data to estimate it empirically), it is always safe to take the worst-case prior probability, as in Eq.~(\ref{eq:worstcasep}), but the latter may result in higher noise. 
Substituting Eq.~(\ref{eqn:p_k}) into~(\ref{eq:maineq}), we can estimate $\epsilon$ value as
\begin{equation}\label{eqn:epsilon_time}
\epsilon_{ij}^k= \frac{- \ln( \frac{P^k_{ij}}{1-P^k_{ij}}(\frac{1}{\delta+ P^k_{ij}} -1))}{r_{ij}}\enspace,
\end{equation}

where $\delta$ is the maximum allowed guessing advantage, and $\epsilon_{ij}^k$ is the DP parameter that can be used to protect the occurrence $k$ of the edge $(i,j)$. To protect \emph{any} possible occurrence in the edge, we assume the worst case by considering the maximum noise, i.e., $\epsilon_{ij}:= min(\epsilon_{ij}^k)$.

To apply a DP mechanism, in addition to $\epsilon_{ij}$, we need to estimate the sensitivity of the query. The sensitivity, in turn, depends on the type of aggregation used to build the DFG. For the sum, min, and max functions, the sensitivity is $\Delta f = 1$, as changing the time duration of one edge occurrence by $1$ time unit will change the output at most by $1$ time unit. The sensitivity of the average aggregation function is $\Delta f = 1/n_{ij}$, where $n_{ij}$ is the total number of occurrences of the edge $(i,j)$. If $R_{ij}$ is the set of all occurrences of the edge $(i,j)$, we can add the noise values drawn from a Laplacian distribution as
\begin{equation*}\label{eqn:m_f}
M_f(R_{ij}) :=f(R_{ij}) + Lap(\frac{\Delta f}{\epsilon_{ij}})\enspace.
\end{equation*}
To measure the effect of injecting the noise on the utility of the values, we use the APE, MAPE, SMAPE as mentioned above in Definition~\ref{def:mape}.

\begin{figure*}
	\begin{subfigure}[b]{0.9\textwidth}
		\centering
		\includegraphics[width=.98\columnwidth]{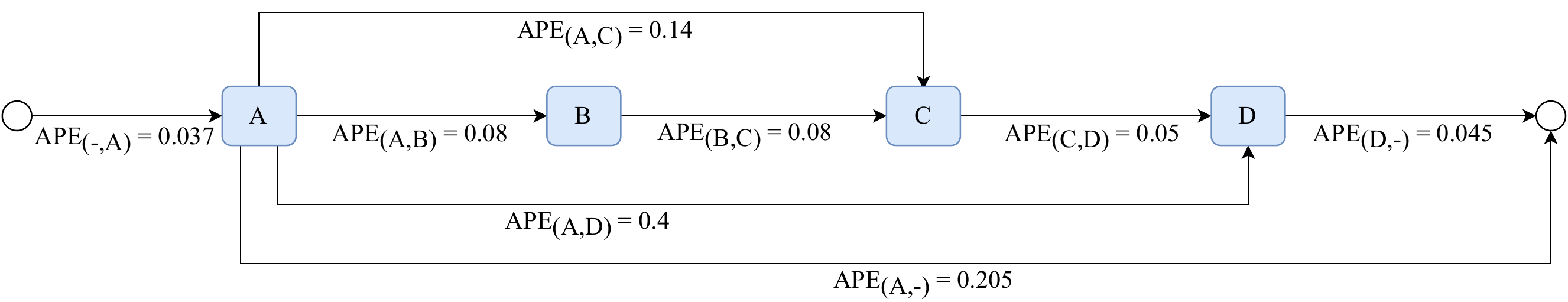}
		\caption{Frequency-annotated Directly-Follows Graph, with an estimated $\epsilon$ =1.695}
		\label{fig:dfg_freq_delta}
		
	\end{subfigure}
	~
	\begin{subfigure}[b]{0.9\textwidth}
		\centering
		\includegraphics[width=.98\columnwidth]{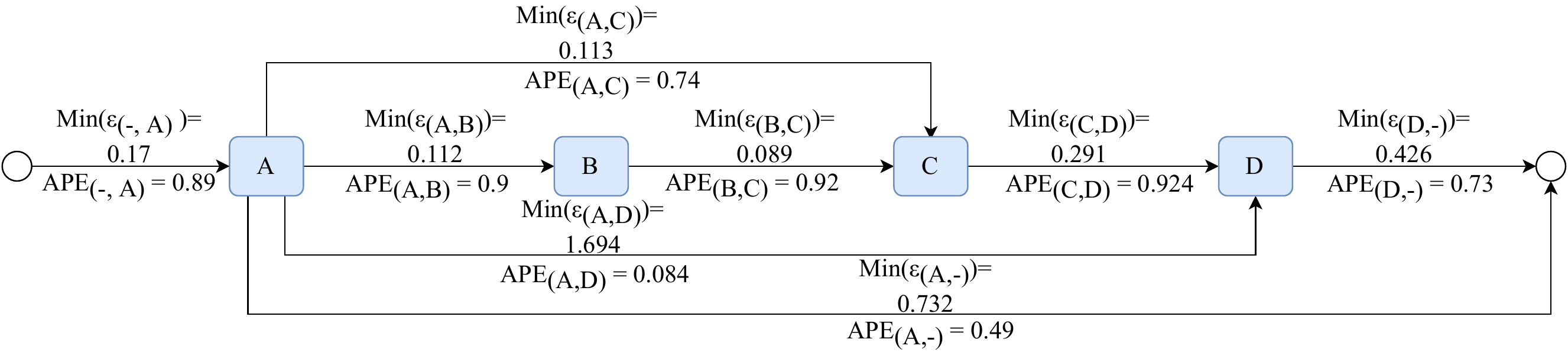}
		\caption{Timely-annotate Directly-Follows Graph with Maximum Aggregation Function}
		\label{fig:dfg_time_delta}
		
	\end{subfigure}
	\caption{ $\epsilon$ and $APE$ with input $\delta$=0.4, for the DFG in Figure~\ref{fig:dfg} }
	\label{fig:running_example_delta}
\end{figure*}

As an example, suppose that a DFG publisher intends to publish the time-annotated DFG in Figure~\ref{fig:dfg_time}. The publisher sets the guessing precision $p=0.1$.
The publisher wants to make sure that the success guessing probability of the analyst to an individual will not increase by more than 40\%, i.e., $\delta=0.4$. To use an $\epsilon$-differential privacy mechanism, the publisher needs to estimate the value of $\epsilon$ for noise generation.  We list the time differences of the input event log in table~\ref{tbl:time_difference}. 
For instance, the directly-follows relation (A, C) as an example has the time differences $1$, $6$, and $15$. That is, $r_{A,C} = max(1,6,15) = 15$. First, we estimate prior guessing probability, within precision $p=0.1$ (i.e., $\pm 1.5$~hours), for every occurrence using Eq~(\ref{eqn:p_k}), $P^1_{A,C} = P^6_{A,C} = P^{15}_{A,C}=\frac{1}{3}$. 
Given the prior knowledge and the guessing advantage, we can estimate an optimal value of $\epsilon$ using Eq~(\ref{eqn:epsilon_time}). The minimum $\epsilon$ over the edge (which is in our case the same for all of them) is $\epsilon_{A,C}$=0.114. We use $\epsilon_{A,C}$ to generate a noise drawn from a Laplace distribution. Suppose that the sampled noise equals 11.156 hours. From Definition~\ref{def:mape}, the $APE$ is 74.3\% to publish the time-annotated DFG with a guessing advantage of $\delta=0.4$. Figure~\ref{fig:dfg_time_delta} shows the $\epsilon$ and $APE$ of every edge in the DFG with $\delta=0.4$ and $p=0.1$. The MAPE of disclosing the time annotated DFG is 0.801.

\begin{table}[hbtp]
	\centering	
	\caption{Time Differences the event log in Table~\ref{tbl:event_log} }
	\begin{tabular}[t] { p{2cm}	p{5cm}	} 
		\hline
		
		Relation     &     Time Differences (hours)       \\\hline 
		(--,A)    &    [1, 1, 1, 2, 2, 2, 3, 3, 4, 5, 10]     \\

		(A,B)    &    [0.2, 3, 8, 12, 16]       \\

		(A,C)    &   [1, 6, 15]        \\

		(A,D)    &    [7]         \\

		(B,C)    &    [1, 5, 11, 15, 20]         \\

		(C,D)    &    [0.2, 0.25, 0.4, 1.5, 2.6, 3.65, 4.7, 6]         \\


		(D,--)    &    [0.5, 0.7, 0.9, 0.9, 0.9, 1.3, 1.3, 1.7, 4]            \\

		(A,--)    &    [0.5, 3]             \\ \hline 
		
	\end{tabular}
	
	\label{tbl:time_difference}
\end{table}

\paragraph{Frequency Annotated DFG.}
The frequency annotated DFG is a count query, which has a single value for each pair, e.g., the number of times an execution of two activities has been observed. In such a case, we cannot estimate the CDF. Therefore,  we take the worst case prior probability as in Eq~(\ref{eq:worstcasep}). The prior knowledge will be the same for all the edges and equals:


\begin{equation}\label{eqn:p_freq}
P=\frac{1-\delta}{2}\enspace.
\end{equation}
Since the attacker is guessing a value $\in \{0,1\}$ (i.e., whether a certain individual participated in the directly-follows relation or not), the value of $r_{ij}$, which is the maximum difference between two possible guesses, equals to $1$. Substituting $r_{ij}=1$ into~(\ref{eq:maineq}), we can calculate the $\epsilon$ for the frequency annotated DFG as 
\begin{equation}\label{eqn:epsilon_freq}
\epsilon = \frac{- \ln( \frac{P}{1-P}(\frac{1}{\delta + P} -1))}{1}\enspace.
\end{equation}
Hence, $P$ and $r_{ij}$ are fixed across the edges, $\epsilon$ will be the same for all the edges. To apply a DP mechanism, we need the sensitivity of the query, which is $\Delta f = 1$, as adding/removing an occurrence from a frequency count changes the output by $1$. If $R_{ij}$ is the set of all occurrences of the edge $(i,j)$, we can add the noise values drawn from a Laplacian distribution as
\begin{equation*}
M_f(R_{ij}) :=f(R_{ij}) + Lap(\frac{1}{\epsilon})\enspace.
\end{equation*}

To measure the effect of injecting noise on the utility of the frequency-annotated DFG, we use percentage error measures mentioned in Definition~\ref{def:mape}.

For example, suppose that a DFG publisher intends to disclose the frequency-annotated DFG in Figure~\ref{fig:dfg_graph}. He needs to make sure that the success probability of an analyst to guess whether an individual participated in a directly-follows relation or not, will not increase by more than 40\% after the DFG disclosure, i.e., $\delta=0.4$. To use an $\epsilon$-differential privacy mechanism, he needs to estimate the value
of $\epsilon$ to generate the noise. For instance, the directly-follows relation (A, C) has a frequency of 3. We estimate the worst-case prior knowledge probability using Eq~(\ref{eqn:p_freq}), $P= 0.3$. After that, we can estimate $\epsilon$ to minimize the guessing advantage to be below $\delta$ from Eq~(\ref{eqn:epsilon_freq}), $\epsilon=1.695$. We then use $\epsilon$ to generate a noise drawn from a Laplace distribution. The noise equals to $0.41$ . The average percentage error APE is $13.5\%$ to publish the noisified directly-follows relation (A, C) with a guessing advantage of $\delta=0.4$. Figure~\ref{fig:dfg_freq_delta} shows the $APE$ of every edge in the frequency annotated DFG with $\delta=0.4$. The MAPE of disclosing the entire frequency annotated DFG is 0.13.

\subsection{P2: From MAPE to $\epsilon$ and Risk}

\begin{figure}[!htb]
	\centering
	\includegraphics[width=.5\columnwidth]{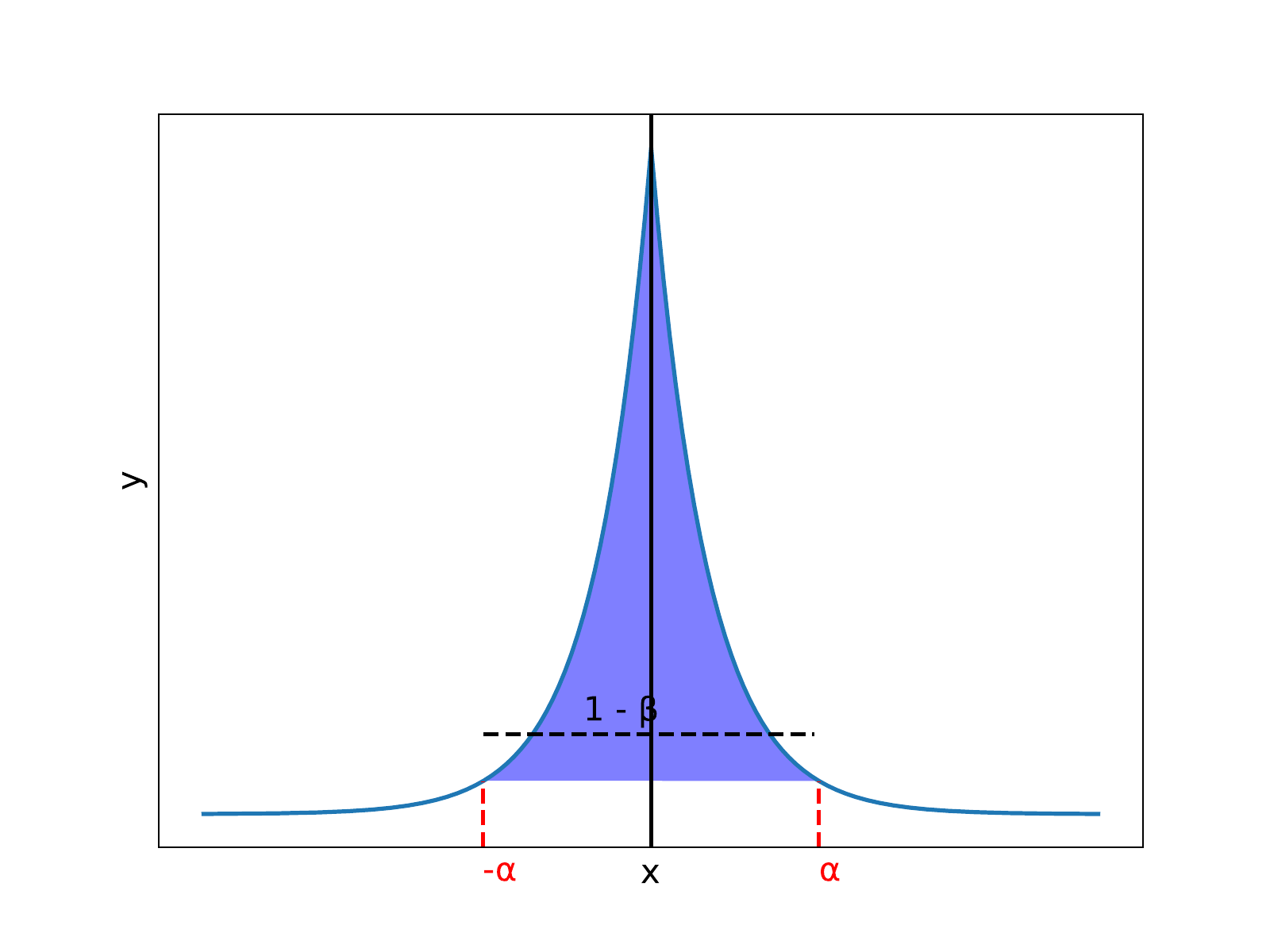}
	\caption{Laplace Noise Distribution, $1-\beta$ is the probability that the added noise to the value $x$ will be within the upper bound $\alpha$ }
	\label{fig:laplace}
\end{figure}

Following, we present the mathematical model of Problem~\ref{int:prob:1},
mapping an input maximum acceptable percentage error to both the amount of noise, $\epsilon$, and the risk measure (guessing advantage). In this article, we adopt the Laplace mechanism.  The $\epsilon$ of the Laplace mechanism can be calculated by observing the distribution of laplace function with scaling $\frac{\Delta f}{\epsilon}$ (e.g~\cite[Fact 3.7]{dwork2014algorithmic}) as follows:
 
\begin{equation}\label{eqn:epsilon_laplace}
\epsilon= \frac{\Delta f}{\alpha}\ln( \frac{1}{\beta} )\enspace,
\end{equation}

\noindent where $\beta$ is the probability that the added noise will be larger than $\alpha$, as shown in Figure~\ref{fig:laplace}. The difference between the true output of the aggregation function of a directly-follows relation   $A_{ij}$ and the noisified value $F_{ij}$ will be at most $\alpha$, with probability $1-\beta$. In this article, we fix $\beta=0.05$ to be statistically insignificant. We assume that the desired upper bound absolute percentage error $\frac{|A_{ij} - F_{ij}|}{|A_{ij}|}$ is the same for all edges. From Definition~\ref{df:MAPE} we can calculate the upper bound of noise for every edge $\alpha_{ij}$ as:
 
\begin{equation} \label{eqn:alpha_per_edge}
\alpha_{ij} = A_{ij}\times MAPE. 
\end{equation} 

Given $\alpha_{ij}$, we can estimate the, $\epsilon$-differential privacy parameter across the edges, $\epsilon_{ij}$ using Eq.~(\ref{eqn:epsilon_laplace}) as follows:
\begin{equation}\label{eqn:epsilon_laplace_per_edge}
\epsilon_{ij}= \frac{\Delta f}{\alpha_{ij}}\ln( \frac{1}{\beta} )\enspace,
\end{equation}

\paragraph{Time Annotated DFG.}
For the time annotated DFG, the directly-follows relations result from of aggregation functions, e.g. maximum over activities duration. Therefore, we have a distribution of event durations for every relation. The guessing advantage of disclosing an event duration, $\delta_{ij}^k$, can be estimated from solving Eq.~(\ref{eqn:epsilon_laplace_per_edge}), and Eq.~(\ref{eqn:epsilon_time}).

%
%
%

\begin{prop}[Risk of a Directly-Follows Relation Occurrence]
	The guessing advantage attached with an occurrence of a directly-follows relation in a time annotated DFG is
	
	$\delta_{ij}^k = \frac{P^k_{ij}} {(1-P^k_{ij})exp(-\epsilon_{ij}\cdot r_{ij}) + P^k_{ij}}-P^k_{ij}$
	
\end{prop}
\begin{proof}
 From Eq.~(\ref{eqn:epsilon_time}), 
we can solve the equation to become:
\begin{eqnarray}\label{eq:delta_time}
\ln( \frac{P^k_{ij}}{1-P^k_{ij}}(\frac{1}{\delta_{ij}^k + P^k_{ij}} -1))&=&-\epsilon_{ij}\cdot r_{ij} \nonumber\\
\text{taking the exponential of both sides,}\nonumber\\
\frac{P^k_{ij}}{1-P^k_{ij}} (\frac{1}{\delta_{ij}^k + P^k_{ij}}-1)&=&exp(-\epsilon_{ij}\cdot r_{ij}) \nonumber\\
\text{multiply  both sides by } \frac{1-P^k_{ij}}{P^k_{ij}}, \nonumber\\
\frac{1}{\delta_{ij}^k + P^k_{ij}}-1&=&\frac{1-P^k_{ij}}{P^k_{ij}} exp(-\epsilon_{ij}\cdot r_{ij}) \nonumber\\
\frac{1}{\delta_{ij}^k + P^k_{ij}}&=&\frac{(1-P^k_{ij})exp(-\epsilon_{ij}\cdot r_{ij}) + P^k_{ij}}{P^k_{ij}} \nonumber\\
\delta_{ij}^k + P^k_{ij}&=&\frac{P^k_{ij}} {(1-P^k_{ij})exp(-\epsilon_{ij}\cdot r_{ij}) + P^k_{ij} } \nonumber\\
\delta_{ij}^k&\!=\!&\frac{P^k_{ij}} {(1\!-\!P^k_{ij})exp(-\epsilon_{ij}\!\cdot\! r_{ij})\!+\!P^k_{ij}}\!-\!P^k_{ij}\enspace.
\end{eqnarray}
\end{proof}

Since each $\delta_{ij}^k$ characterizes the risk of a particular occurrence of the set of the directly-follows relation (i, j), we can estimate the overall guessing advantage of a directly-follows relation, $\delta_{ij}$.
\begin{definition}[Risk of a Directly-Follows Relation Disclosure]

	The guessing advantage of a time annotated directly-follows relation between two activities is the maximum guessing advantage of all occurrences of that relation. 
\begin{equation}\label{eq:delta_time_2_1}
\delta_{ij} = \max_{k}(\delta_{ij}^k)\enspace.
\end{equation}
\end{definition}

Finally, from the Eqs.~(\ref{eq:delta_time}), and ~(\ref{eq:delta_time_2_1}), we can calculate the guessing advantage of disclosing the entire DFG, $\delta$. 

\begin{definition}[Risk of a DFG Disclosure]

	The guessing advantage of a directly-follows graph is the maximum guessing advantage of its directly-follows relations. 
	\begin{equation}\label{eq:delta_time_2}
	\delta = \max_{i,j}(\delta_{ij})\enspace.
	\end{equation}
\end{definition}

\begin{table}[hbtp]
	\centering	
	\caption{Estimated risk measure (guessing advantage) of time differences of the event log in Table~\ref{tbl:event_log} with MAPE=0.3, for the Max aggregation function }
	\begin{tabular}[t] { p{2cm}	p{5cm}	} 
		\hline
		
		Relation     &    Estimated Risk Measure ($\delta_{ij}^k$) per occurrence      \\\hline 
		(--,A)    &    [0.636, 0.636, 0.636, 0.455, 0.455, 0.455, 0.545, 0.545, 0.727, 0.909, 0.909]     \\

		(A,B)    &    [0.799, 0.799, 0.799, 0.799, 0.799]       \\

		(A,C)    &   [0.667, 0.667, 0.667]        \\

		(A,D)    &    [0.342]         \\

		(B,C)    &   [0.799, 0.799, 0.799, 0.799, 0.799]         \\

		(C,D)    &    [0.625, 0.625, 0.625, 0.875, 0.875, 0.875, 0.875, 0.875]         \\


		(D,--)    &    [0.667, 0.444, 0.444, 0.444, 0.444, 0.444, 0.444, 0.667, 0.889]            \\

		(A,--)    &    [0.499, 0.499]            \\ \hline 
		
	\end{tabular}
	
	\label{tbl:time_risk}
\end{table}

For example, suppose that a DFG publisher intends to publish the time-annotated DFG in Figure~\ref{fig:dfg_time}. The publisher chooses the guessing precision $p=0.1$,
 and he needs the average percentage error of every edge  to be within $30\%$, with a probability $1-\beta$ (we assumed $\beta=0.05$ as mentioned above), i.e., MAPE=$0.3$. The publisher aims to use an $\epsilon$-differential privacy mechanism due to its composability. He needs to estimate the value of $\epsilon$ to generate the noise. 
Table~\ref{tbl:time_difference} list the time differences over the edges.

\begin{figure*}
	\begin{subfigure}[b]{0.9\textwidth}
		\centering
		\includegraphics[width=.98\columnwidth]{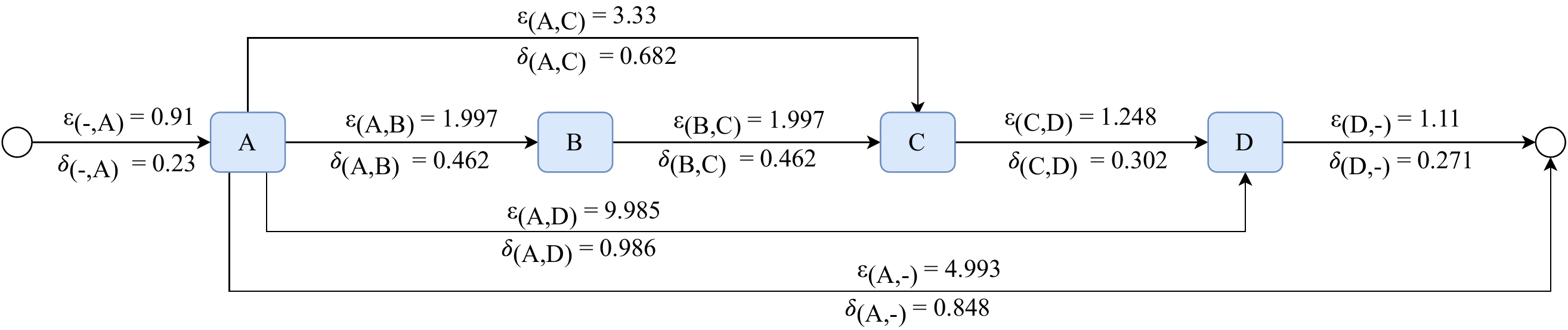}
		\caption{Frequency-annotated Directly-Follows Graph}
		\label{fig:dfg_freq_mape}
		
	\end{subfigure}
	
	~
	\begin{subfigure}[b]{0.9\textwidth}
		\centering
		\includegraphics[width=.98\columnwidth]{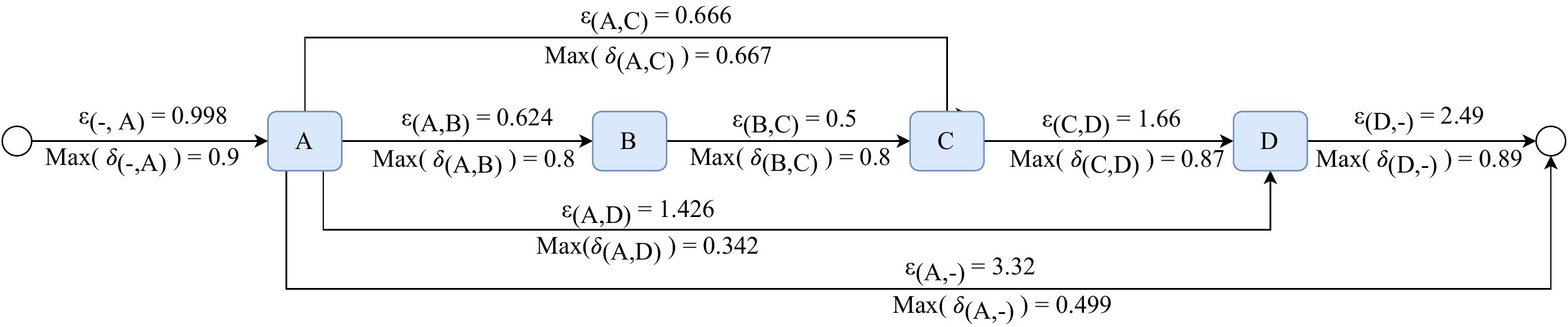}
		\caption{Timely-annotate Directly-Follows Graph with Maximum Aggregation}
		\label{fig:dfg_time_mape}
		
	\end{subfigure}
	\caption{ $\epsilon$ and $\delta$ calculations with input $MAPE$=0.3, for the DFG in Figure~\ref{fig:dfg} }
	\label{fig:running_example_mape}
\end{figure*}

For instance, the time differences of the directly-follows relation (A, C) are $1$, $6$, and~$15$. From Eq~(\ref{eqn:alpha_per_edge}), $\alpha_{A, C}=15 \cdot 0.3= 4.5$. From Eq~(\ref{eqn:epsilon_laplace_per_edge}), $\epsilon_{A, C}=0.667 $. And lastly, from Eq(~\ref{eq:delta_time}), we take the worst-case, i.e., the maximum $\delta$, so $\delta_{A, C}^1=0.666$, $\delta_{A, C}^6=0.666$ and $\delta_{A, C}^{15}=0.666$, and from Eq~(\ref{eq:delta_time_2_1}), we can estimate $\delta_{A, C}=0.666$. The  $\delta_{ij}^k$ of the time differences of all the edges are given in Table~\ref{tbl:time_risk}, and Figure~\ref{fig:dfg_time_mape} shows both the $\epsilon_{ij}$ and $\delta_{ij}$ of the of every edge of the DFG. From Eq~(\ref{eq:delta_time_2}), we take the maximum guessing advantage  of all the edges to be the guessing advantage of disclosing the entire DFG. The guessing advantage of publishing the time annotated DFG is $\delta= 0.9$ to keep the APE$\leq 30\%$.

\paragraph{Frequency Annotated DFG.}
For the frequency annotated DFG, the weights of the edges are the number of occurrences of the directly-follows relations. The guessing advantage of a frequency annotated directly-follows relation, $\delta_{ij}$, can be estimated by solving Eqs.~(\ref{eqn:p_freq}) and~(\ref{eqn:epsilon_time}).

\begin{prop}[Risk of frequency-annotated directly-follows relation]
	The guessing advantage of a frequency annotated directly-follows relation between two activities is 
	\begin{equation}
	\delta_{ij} = \frac{1-\sqrt{\exp(-\epsilon_{ij})}}{1+\sqrt{\exp(-\epsilon_{ij})}}	
	\end{equation}

\end{prop}
\begin{proof}
	Substituting $P^k_{ij} := P$ from~(\ref{eqn:p_freq}) into~(\ref{eqn:epsilon_time}), and for the frequency, as discussed above, $r_{ij}=1$, hence we get
	\begin{eqnarray}\label{eq:delta_freq}
	\frac{P}{1-P}(\frac{1}{\delta_{ij} + P} -1)&=&\exp(-\epsilon_{ij}) \nonumber\\
	 \text{substituting } P=\frac{1-\delta}{2}, \nonumber \\
	\frac{1-\delta_{ij}}{1+\delta_{ij}}(\frac{1}{\delta_{ij} + \frac{1-\delta_{ij}}{2}} -1)&=&\exp(-\epsilon_{ij}) \nonumber\\
	\frac{1-\delta_{ij}}{1+\delta_{ij}}(\frac{1}{\frac{1+\delta_{ij}}{2}}-1 )&=&\exp(-\epsilon_{ij}) \nonumber\\
	\frac{1-\delta_{ij}}{1+\delta_{ij}}(\frac{1-\delta_{ij}}{1+\delta_{ij}} )&=&\exp(-\epsilon_{ij}) \nonumber\\
		\text{taking the square root of both sides, }\nonumber \\
	\frac{1-\delta_{ij}}{1+\delta_{ij}}&=&\sqrt{\exp(-\epsilon_{ij})} \nonumber\\
	1-\delta_{ij}&=&\sqrt{\exp(-\epsilon_{ij})}+\delta_{ij}\cdot\sqrt{\exp(-\epsilon_{ij})} \nonumber\\
	1-\sqrt{\exp(-\epsilon_{ij})}&=&\delta_{ij}+\delta_{ij}\cdot\sqrt{\exp(-\epsilon_{ij})} \nonumber\\
	1-\sqrt{\exp(-\epsilon_{ij})}&=&\delta_{ij}\cdot(1+\sqrt{\exp(-\epsilon_{ij})}) \nonumber\\
\text{multiply both sides by } \frac{1}{1+\sqrt{\exp(-\epsilon_{ij})}},\nonumber\\
	\delta_{ij} &=& \frac{1-\sqrt{\exp(-\epsilon_{ij})}}{1+\sqrt{\exp(-\epsilon_{ij})}} \enspace.
	\end{eqnarray}
	
	\end{proof}

\begin{definition}[Risk of a Frequency Annotated DFG Disclosure]
	
	The guessing advantage of a frequency annotated directly-follows graph is the maximum guessing advantage of the directly-follows relations. 
	\begin{equation}\label{eq:delta_freq_2}
	\delta = \max_{i,j}(\delta_{ij})\enspace.
	\end{equation}
\end{definition}

For example, a DFG publisher intends to disclose the frequency-annotated DFG in Figure~\ref{fig:dfg_graph}. He wants to make sure that the average percentage error does not exceed 30\%, i.e., $MAPE= 0.3$. He aims to use an $\epsilon$-differential privacy mechanism to generate the noise. 
For instance, the directly-follows relation (A, C) has a frequency of 3. First, we need to estimate $\alpha_{A, C}$ from Eq(~\ref{eqn:alpha_per_edge}), $\alpha_{A, C}=0.899$. Then, we can estimate $\epsilon_{A, C}$ from Eq(~\ref{eqn:epsilon_laplace_per_edge}), $\epsilon_{A, C}=3.329$. We can estimate the $\delta_{A, C}$ from Eq(~\ref{eq:delta_freq}), $\delta_{A, C}= 0.682$. Figure~\ref{fig:dfg_freq_mape} shows the $\epsilon_{ij}$ and $\delta_{ij}$ values across the frequency-annotated DFG. The guessing advantage of publishing the entire frequency annotated DFG is $\delta=0.986$. In this example, the $\delta$ is close to 1 because the relation (A, D) has only one occurrence, which makes $\delta_{A, D}= 0.986$, and it is difficult to protect a single value with a 30\% absolute error.

\section{Evaluation}
\label{sec:evaluation}


We evaluate the proposed model by studying the relation between its parameters: $\delta$, $\epsilon$, and MAPE to address the problems~\ref{int:prob:1} and~\ref{int:prob:1} . We study the following research questions:

\begin{itemize}
\item RQ1: What is the effect of the risk control parameter $\delta$ (guessing advantage), on the process model utility loss (average percentage error)?

\item RQ2: What is the effect of maintaining a utility loss level (average percentage error) on the risk control parameter $\delta$?
\end{itemize}
We evaluate the protected DFGs by comparing them to ground truth. We use the unprotected DFG, without any applied privacy mechanism, as our ground truth. We compare both the results of frequency and time directly-follows graphs. Also, we report the results of more than one aggregate type. We study the distributions of model parameters across the edges of the DFG. We use 13 real-life event logs for a better understanding of the relation between the model parameters. The exact values of all the reported distributions are available in the supplementary material~\cite{supplementaryMaterial}. We start with the experimental set-up.

\subsection{Experimental Set-up}
For our experiment, we use the publicly available set of real-life event logs at 4TU Centre for Research Data\footnote{https://data.4tu.nl/} as of October 2020. We considered the event logs mentioned in Table~\ref{tab:event_logs}. The listed files log the process execution from different domains, e.g. finance, government, healthcare, and IT. In our experiment, we excluded the event logs that do not record explicit business processes (i.e., ``Logdata of interactions with three interfaces", ``NASA CEV", ``BPIC 2016" logs, ``Activities of daily living of several individuals", ``Apache Commons Crypto 1.0.0", ``Statechard workbench and Aligment", ``Junit 4.12", and ``Data underlying the paper: Automated Discovery..."). Also, we excluded the event logs ``coSeLog WABO" as they are the original pre-preprocessed version of BPIC 2015. We excluded the BPIC 2018 from our experiment as its execution time is longer than 10 hours. Finally, in the BPIC 2013, BPIC 2014, BPIC 2015, BPIC 2017, and BPIC 2020 we selected only one event log from every group. The selected BPIC files are listed in Table~\ref{tab:event_logs}.

We report the characteristics and the descriptive statistics of the event logs in Table~\ref{tab:event_logs}. The listed event logs are heterogeneous. There are small and simple event logs, e.g., CCC 2019, which contains only 20 traces and 20 case variants. There are large and complex event logs, e.g., BPIC 2019, which contains 251734 cases with 11973 case variants. Some of the event logs contain challenging properties, e.g., BPIC 2015, which contains 1170 case variants although the number of cases is 1199, and Sepsis event log, which contains 846 case variants and only 1050 cases. Also, some of the event logs contain outliers in the distribution of time differences, e.g. the average case duration of BPIC 2015 is 3.15 months, and the maximum case duration is 4.07 years, which adds to the uniqueness of individual cases. These sources of uniqueness make the re-identification risk higher. We also included the Credit Requirement event log, which contains 10035 cases, with only one case variant, which means all the cases are following the same process.

\begin{table}[hbtp]
	\centering
	\caption{Descriptive Statistics of Event Logs}
	\label{tab:event_logs}
	\footnotesize
	\begin{tabular}{|c|c|c|c|c|c|c|c|c|c|c|c|}
		\hline
		
		\multirow{2}{*}{event log}
		&\multirow{2}{3em}{\centering\# Traces}
		&\multirow{2}{3em}{\centering\# Tasks}
		&\multirow{2}{3em}{\centering\# Events}
		&\multirow{2}{3em}{\centering\# Edges}
		&\multirow{2}{3em}{\centering Case Variant}
		&\multicolumn{2}{c|}{Trace Length}
		& \multicolumn{3}{c|}{Case Duration}\\
		
		\cline{7-11}
		
		& & & & & &  Min&Max&Min &Max &Avg\\

		\hline
		
		$BPI12$~\cite{BPIC2012}	&	13087	&	23	&	262200	&	116	&	4366	&	3	&	175	&	1.85 s	&	4.51 m	&	1.23 w	\\
				\hline
		$BPI13_{i}$~\cite{BPIC13}	&	7554	&	4	&	65533	&	16	&	1511	&	1	&	123	&	inst.	&	2.11 y	&	1.73 w	\\
				\hline
		$BPI14_{i}$~\cite{BPIC14}	&	46616	&	39	&	466737	&	497	&		22632	&	1	&	178	&	14 s	&	1.07 y	&	5.07 d	\\
				\hline
		$BPI15_1$~\cite{BPIC15}	&	1199	&	398	&	52217	&	495	&	1170	&	2	&	101	&	8.56 h	&	4.07 y	&	3.15 m	\\
				\hline
		$BPI17$~\cite{BPIC2017}	&	31509	&	24	&	1202267	&	181		&	3942	&	10	&	180	&	3.35 m	&	9.4 m	&	3.13 w	\\
				\hline
		$BPIC19$~\cite{BPIC2019}	&	251734	&	42	&	1595923	&	498		&	11973	&	1	&	990	&	2 ms	&	70.33 y	&	2.35 m	\\
				\hline
		$BPI20_{r}$~\cite{BPIC20}	&	7065	&	51	&	86581	&	500	&		1478	&	3	&	90	&	12.61 h	&	3.26 y	&	2.87 m	\\
				\hline
		$CCC19$~\cite{CCC2019}	&	20	&	29	&	1394	&	149	&	20	&	52	&	118	&	11 m	&	1.01 d	&	1.73 h	\\
				\hline
		$CredReq$~\cite{creditReq}	&	10035	&	8	&	150525	&	9		&	1	&	15	&	15	&	3.5 h	&	5 d	&	22 h	\\
				\hline
		$Hospital$~\cite{hospital}	&	1143	&	624	&	150291	&	903		&	981	&	1	&	1814	&	inst.	&	3.17 y	&	1.06 y	\\
				\hline
		$Sepsis$~\cite{felix2017sepsis}	&	1050	&	16	&	15214	&	115		&	846	&	3	&	185	&	2.03 m	&	1 y	&	4 w	\\
				\hline
		$Traffic$~\cite{traffic}	&	150370	&	11	&	561470	&	77		&	231	&	2	&	20	&	3 d	&	12 y	&	11 m	\\
				\hline
		$Unrine.$~\cite{Gunst2020}	&	1650	&	10	&	6973	&	25		&	50	&	2	&	35	&	10.1 m	&	2.32 y	&	3.7 w	\\
				\hline

	\end{tabular}
\end{table}




We implement the proposed model as part of a prototype, namely Amun\footnote{https://github.com/Elkoumy/amun}. In our experiment, we calculated frequency queries by counting the occurrences of each edge. For the time difference queries, we subtract the timestamps of all the occurrences of the edges. The result is a distribution of values over edges. We calculate the four aggregation operations: sum, maximum, minimum, and average for every pair. The sensitivity of the aggregation operations is calculated, as mentioned in Section~\ref{sec:balance}. In our experiment, we choose the time unit (e.g. Minutes, Hours, Days, etc. ) based on the aggregation result. We do that to keep the value between 1 and 1000. Also, we ignore the start and end timestamp of each activity, and we choose the end to start timestamp difference as the directly-follows relation time for simplicity and the same technique could be used to apply differential privacy. To perform the experiment, we use a single machine with Intel(R) core(TM) i5 CPU @ 1.60GHz and 16 GB memory. We use python 3.7 and multiprocessing to leverage multiple cores to parallelize the program execution. We use PM4PY library~\cite{pm4py} for parsing XES files, and we use \textit{diffprivlib} library~\cite{diffprivlib} for drawing noise values from a Laplace distribution.

In the following experiments, we keep the precision parameter $p=0.5$, which means that we consider the attacker's guess successful if he gets as close to the real value as $0.5$ of the maximum value. In a practical application, the user decides the acceptable guessing precision. In the following, we are conducting experiments that evaluate the proposed method to answer the above questions.

\subsection{P1 from $\delta$ to $\epsilon$ and MAPE}
\begin{figure*}[t!]
	\centering
	\begin{subfigure}[b]{0.48\textwidth}
		\centering
		\includegraphics[width=.98\columnwidth]{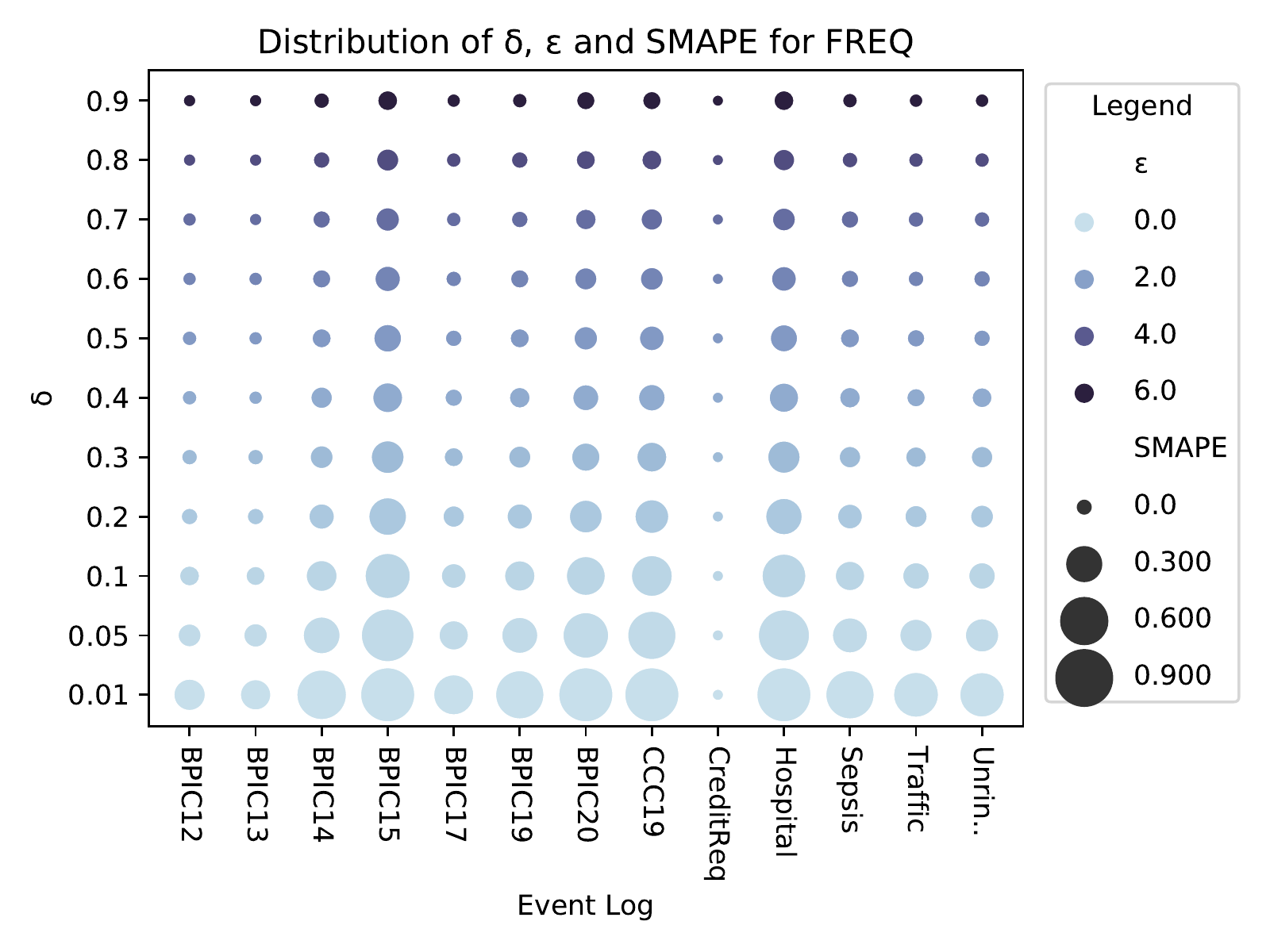}
		\caption{Frequency Queries}
		\label{fig:delta_freq}
	\end{subfigure}\quad
	\begin{subfigure}[b]{0.48\textwidth}
		\centering
		\includegraphics[width=.98\columnwidth]{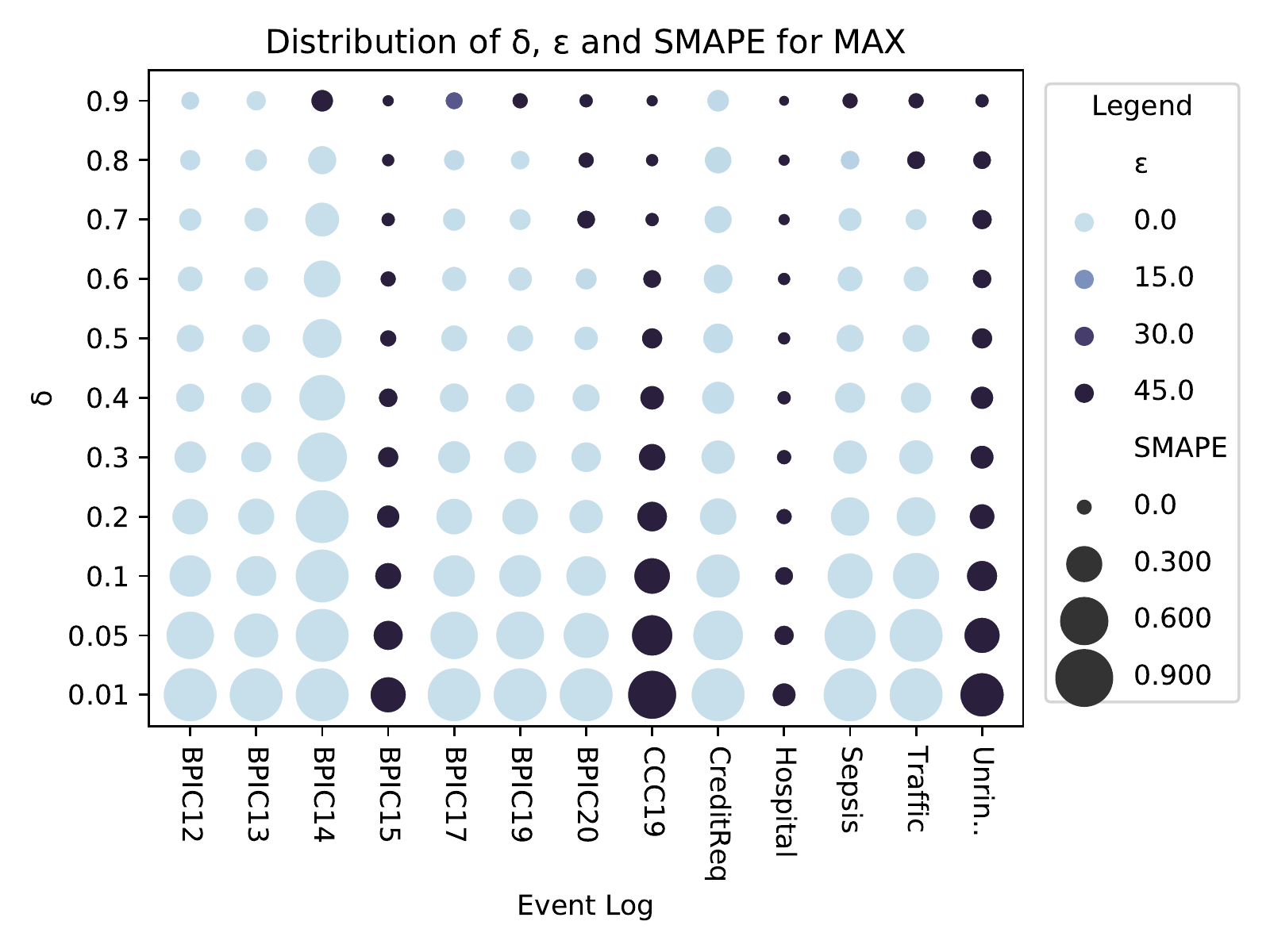}
		\caption{Max Time Difference Queries}
		\label{fig:delta_max}
	\end{subfigure} 
	\bigskip
	\begin{subfigure}[b]{0.48\textwidth}
		\centering
		\includegraphics[width=.98\columnwidth]{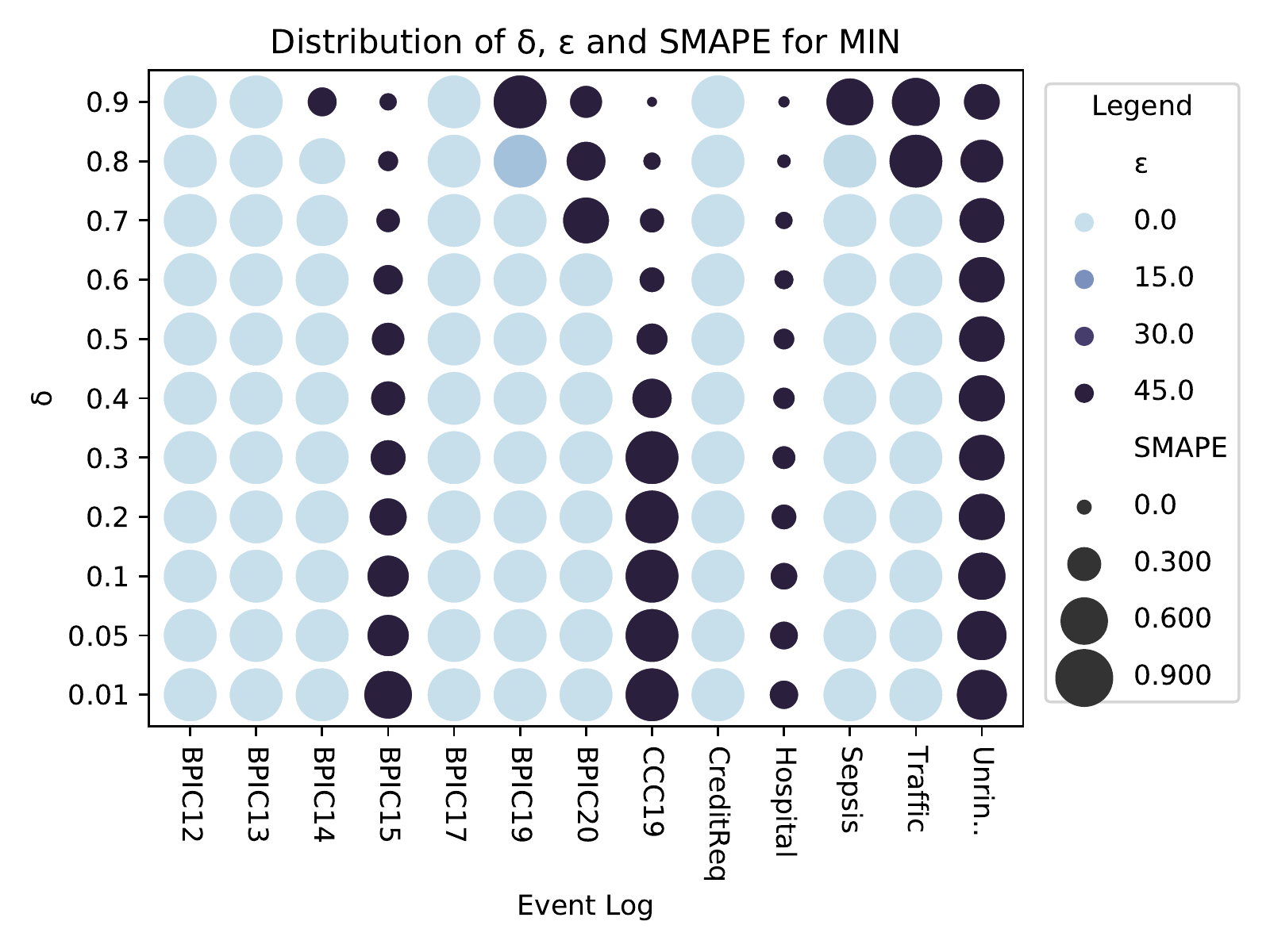}
		\caption{Min Time Difference Queries}
		\label{fig:delta_min}
	\end{subfigure}\quad
	\begin{subfigure}[b]{0.48\textwidth}
		\centering
		\includegraphics[width=.98\columnwidth]{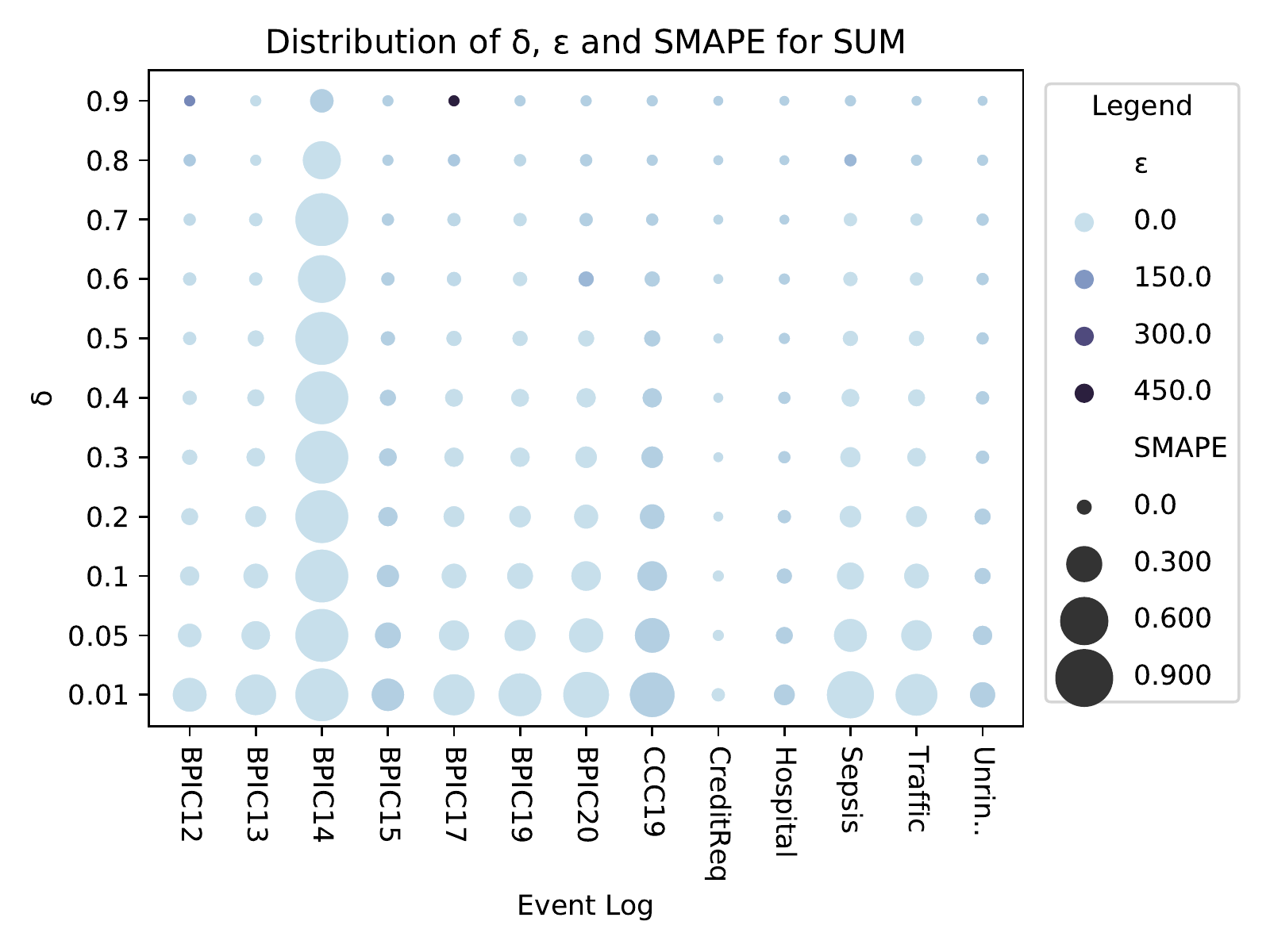}
		\caption{Total Time Difference Queries}
		\label{fig:delta_sum}
	\end{subfigure}
	\bigskip
	\begin{subfigure}[b]{0.48\textwidth}
		\centering
		\includegraphics[width=.98\columnwidth]{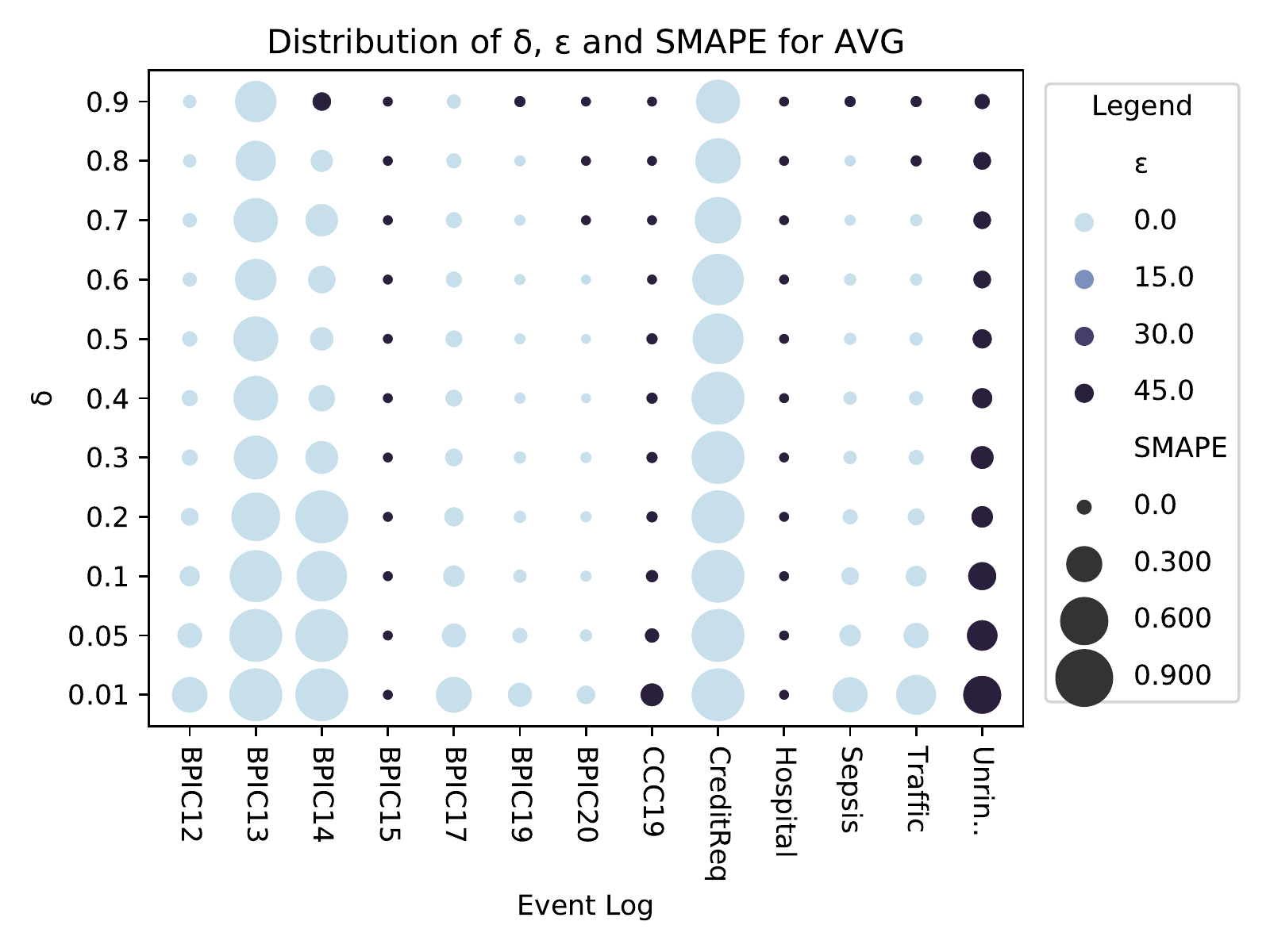}
		\caption{Average Time Difference Queries}
		\label{fig:delta_avg}
	\end{subfigure}
	
	\caption{ $\delta$'s effect on $\epsilon$ and SMAPE. The columns are the event logs, the rows are guessing advantage $\delta$, the size of the circles is the amount of error SMAPE, and The color hue is the  $\epsilon$.}
	\label{fig:bubble_delta}
\end{figure*}

In the setting of problem~\ref{int:prob:1}, a data publisher has a maximum level of risk ($\delta$) that needs to be achieved. The user needs to calculate the differential privacy parameter, $\epsilon$, to achieve that level, and he wants to know how the utility of the discovered model would be after applying the privacy mechanism. We use Eq.~(\ref{eqn:epsilon_freq}) to calculate $\epsilon$ for every edge separately. Then we calculate the amount of noise drawn from the Laplacian distribution with the calculated $\epsilon$. We add the corresponding noise to every edge separately that maintains the given guessing level, $\delta$. To study the effect of  guessing advantage level (risk measure) $\delta$ on the accuracy of the output DFG, we calculate the percentage error over every edge. We use both the MAPE and SMAPE, as mentioned above. We performed the experiments 10 times and calculated the average values across multiple runs. 
We report the effect of $\delta$ on the amount of injected noise (i.e., the differential privacy parameter $\epsilon$) and the percentage error.

In Figure~\ref{fig:bubble_delta}, we report the relation between $\delta$, $\epsilon$ and SMAPE. We use the bubble heat maps, where the columns of every figure represent the event logs, the rows are different values for guessing advantage limit $\delta$. The size of the circles is the percentage error SMAPE. The color hue is the differential-privacy parameter, $\epsilon$, the darker the intensity, the higher $\epsilon$, the lower the amount of noise injected. As we use different $\epsilon$ value across edges, we report the median $\epsilon$ across edges of event logs. The exact values for $\epsilon$, MAPE and SMAPE could be found in the supplementary material as comma-separated values~\cite{supplementaryMaterial}. We log the values for every edge of each event log.

\begin{figure*}[t!]
	\centering
	\begin{subfigure}[b]{0.48\textwidth}
		\centering
		\includegraphics[width=.98\columnwidth]{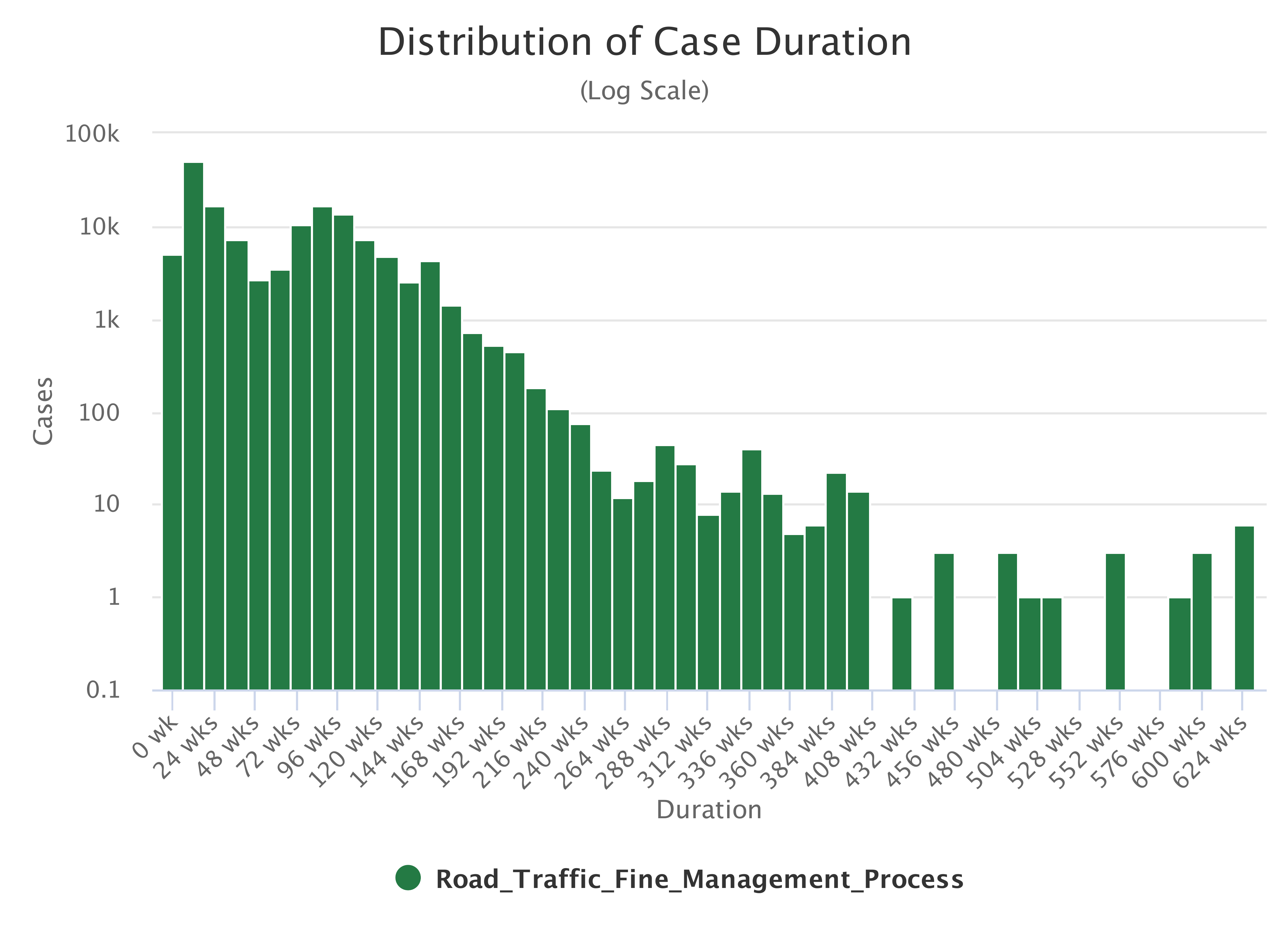}
		\caption{Distribution of Case Duration of Road Traffic Fines (Log Scale)}
		\label{fig:duration_traffic}
	\end{subfigure} \quad
	\begin{subfigure}[b]{0.48\textwidth}
	\centering
	\includegraphics[width=.98\columnwidth]{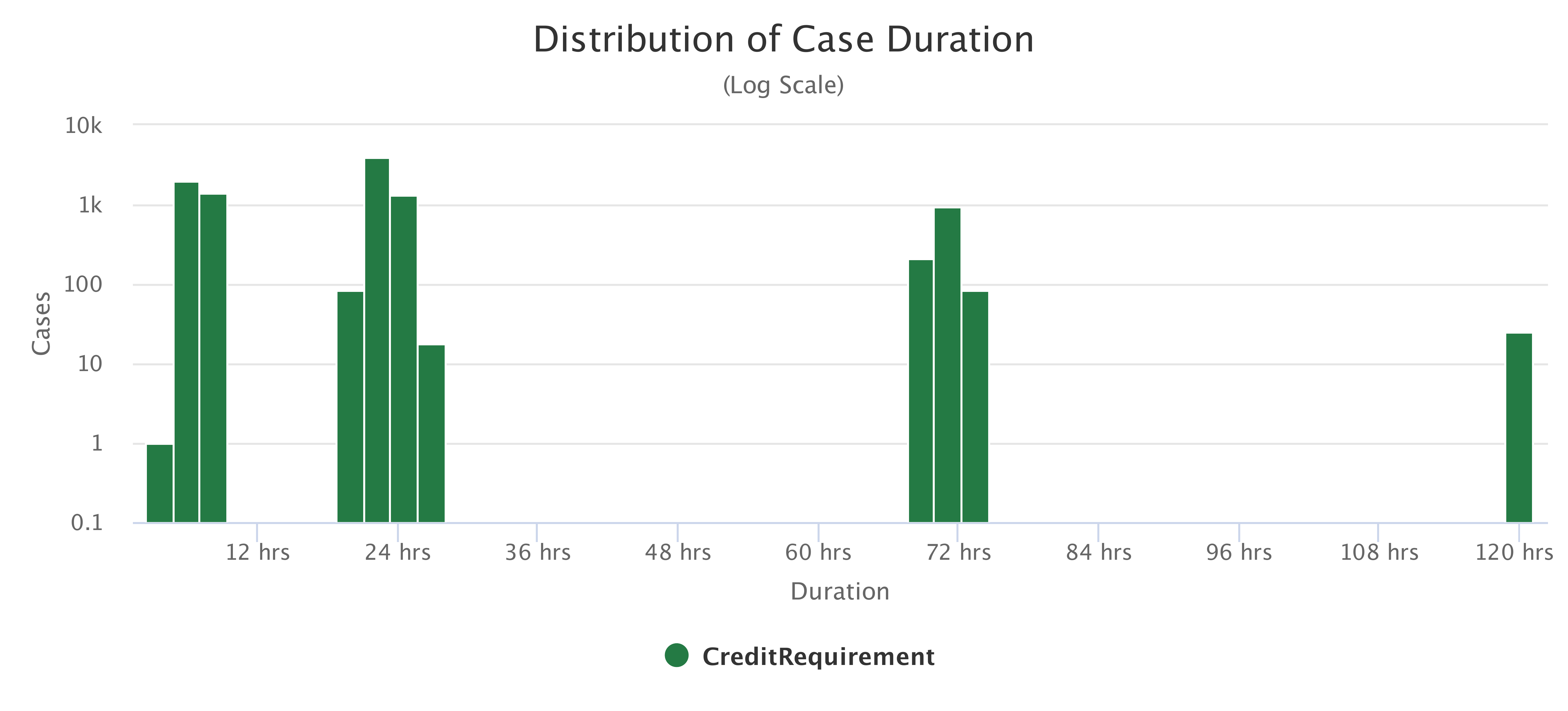}
	\caption{Distribution of Case Duration of Credit Requirement (Log Scale)}
	\label{fig:duration_credit}
\end{subfigure}
	\bigskip
	\begin{subfigure}[b]{0.48\textwidth}
		\centering
		\includegraphics[width=.98\columnwidth]{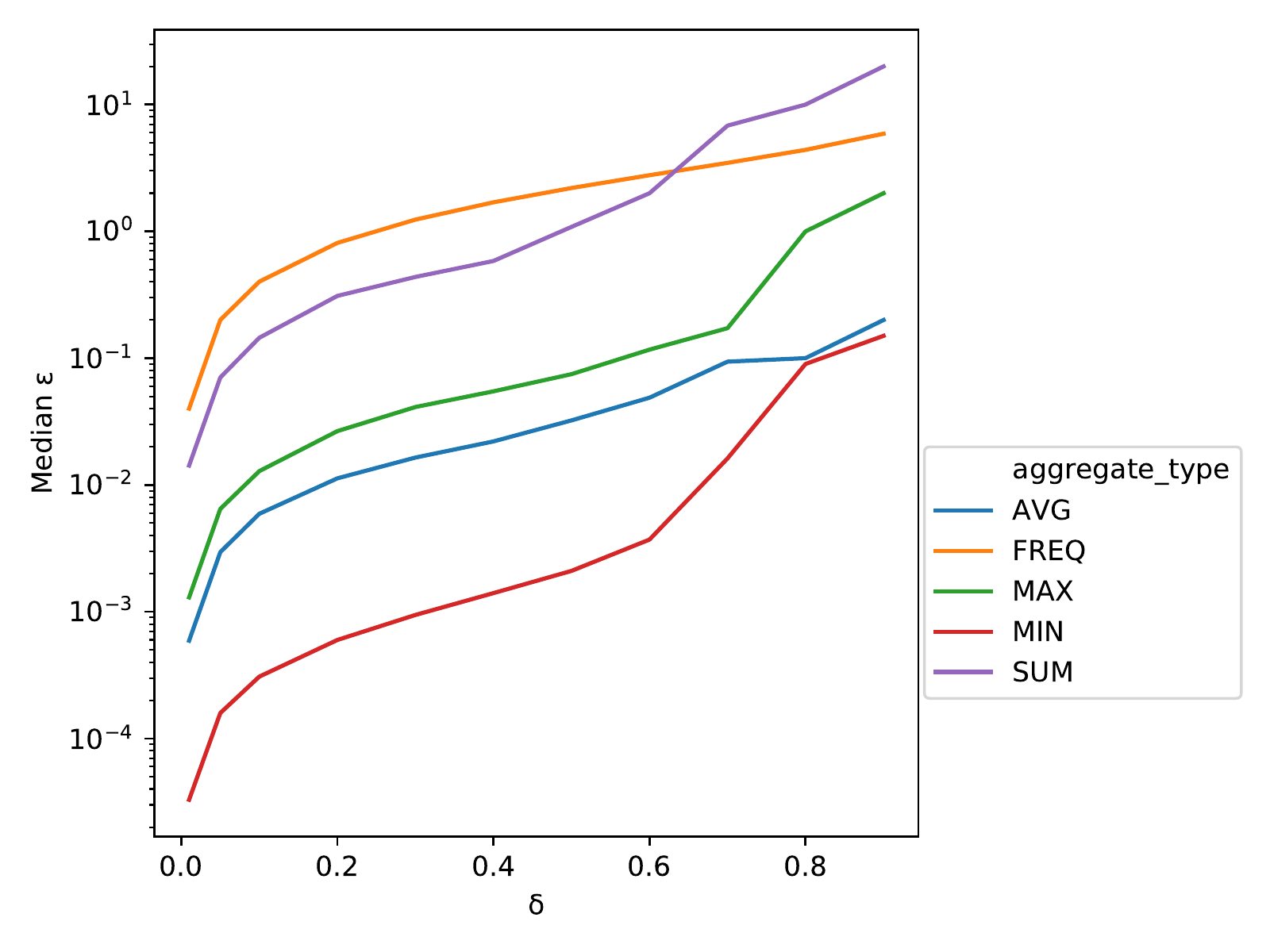}
		\caption{Epsilon Distribution of Road Traffic Fines (Log Scale)}
		\label{fig:epsilon_traffic}
	\end{subfigure}\quad
\begin{subfigure}[b]{0.48\textwidth}
	\centering
	\includegraphics[width=.98\columnwidth]{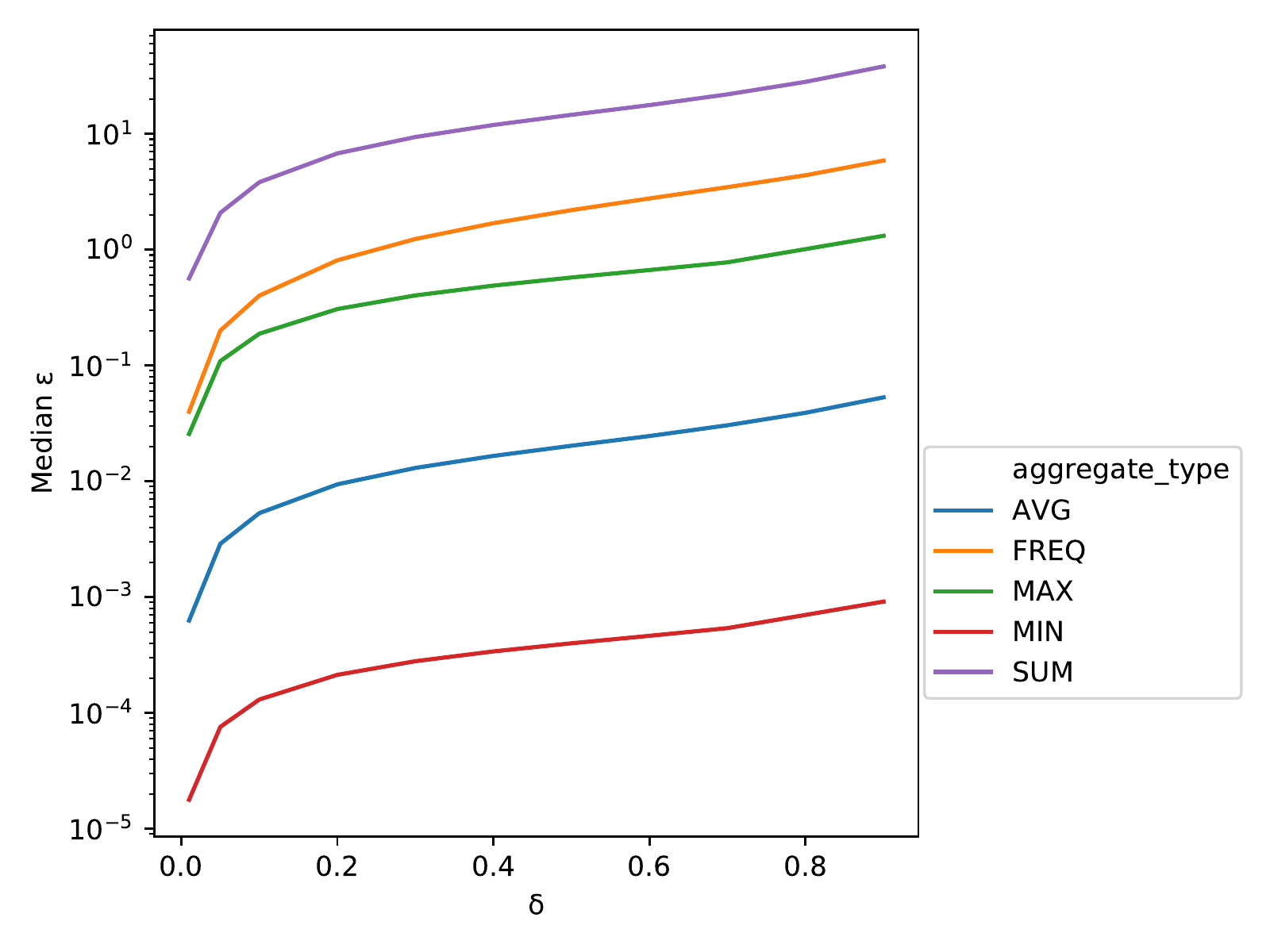}
	\caption{Epsilon Distribution of Credit Requirement (Log Scale)}
	\label{fig:epsilon_credit}
\end{subfigure}
	
	\caption{ Distribution of Case Duration and $\epsilon$ of both Credit Requirement and Traffic Fines event logs}
	\label{fig:distribution_epsilon}
\end{figure*}

In Figure~\ref{fig:bubble_delta}, the value of $\epsilon$ increases with the increase of $\delta$ (the change in color intensity over the column of every event log), because maintaining low-risk values requires a large amount of noise, and hence a smaller $\epsilon$ value.To study the relation of $\epsilon$ and the distribution of the input values, we plot the distribution of $\epsilon$ for both Road Traffic and Credit Requirement in Figure~\ref{fig:distribution_epsilon}. The growth rate of $\epsilon$ differs across the event logs and the aggregation type as it depends on the distribution of values in every edge in every event log and the sensitivity of the aggregation function, as mentioned in Section~\ref{sec:balance}. For instance, in Figure~\ref{fig:delta_max} the Road Traffic event log has the largest noise since the distribution of case duration is right-skewed, and there is a small number of outlier cases. The distribution of case duration and $\epsilon$ of Road Traffic event log are shown in Figures~\ref{fig:duration_traffic} and~\ref{fig:epsilon_traffic}.  On the other hand, the Credit Requirement event log has the maximum $\epsilon$, hence the minimum injected noise for all aggregate types except the minimum because the maximum case duration is 5 days and that happens in 5 cases, which increases the anonymity and the minimum duration is 3 hours, and that happens only in one case. The distribution of case duration and $\epsilon$ for Credit requirement are shown in Figures~\ref{fig:duration_credit} and~\ref{fig:epsilon_credit}. 

The effect of maintaining low-risk levels on the utility appears in Figure~\ref{fig:bubble_delta}. The SMAPE, represented by the size of circles, decreases with the increase of risk. The behavior also changes from one event log to the other as the distribution of values of every event log affects the probability of guessing advantage as explained in Section~\ref{sec:balance} and hence the amount of noise to be added is larger to keep the guessing advantage below $\delta$. For instance, in Figure~\ref{fig:delta_max} the maximum time difference query of the Hospital event log achieves low-risk values with high utility, and that is because the maximum duration is 3 years and there are 17 cases which have 3 years case duration. In most of the event logs, we can achieve 10\% risk with 20\% SMAPE. The distribution of error changes across edges. The exact values of APE are available in the supplementary material~\cite{supplementaryMaterial}. Controlling which aggregation function is permitted and which edges could be revealed, could enhance the balance between risk and utility. We report the distribution of risk over edges in the supplementary material.

In some event logs with frequency annotated DFG, the amount of noise to achieve 1\% risk measure is large, and the utility loss is large, e.g. in Figure~\ref{fig:delta_freq} Hospital and BPIC15 event logs, because both the event logs contain a large number trace variant in comparison to the number of cases , and both contain outliers in the execution time. In some event logs, we can achieve minimal risk limits with minimal utility loss, e.g., BPIC12, Traffic, and Credit Requirement.



\subsection{P2 from MAPE to $\epsilon$ and $\delta$}
\begin{figure*}[t!]
	\centering
	\begin{subfigure}[b]{0.48\textwidth}
		\centering
		\includegraphics[width=.98\columnwidth]{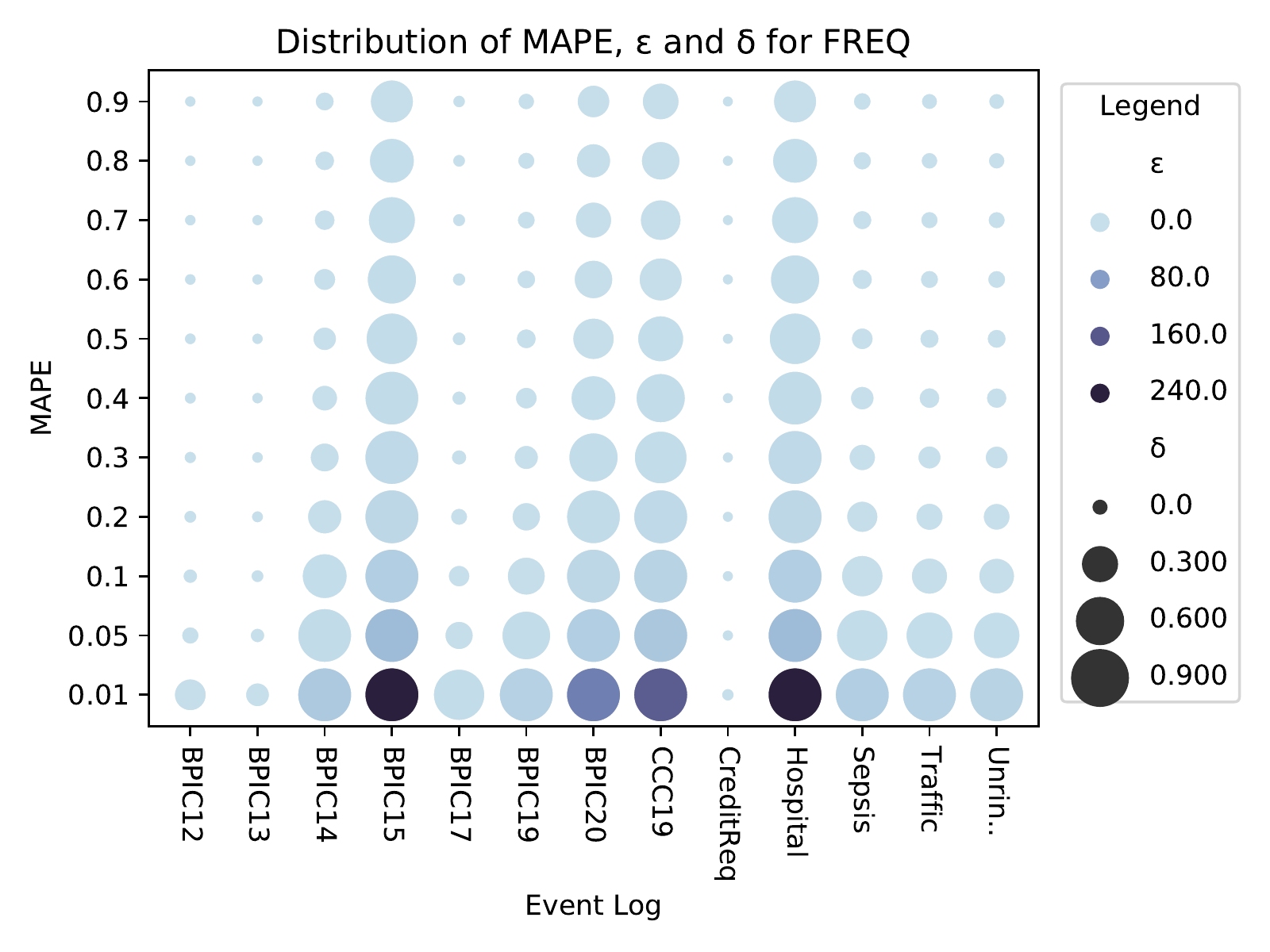}
		\caption{Frequency Queries}
		\label{fig:alpha_freq}
	\end{subfigure}\quad
	\begin{subfigure}[b]{0.48\textwidth}
		\centering
		\includegraphics[width=.98\columnwidth]{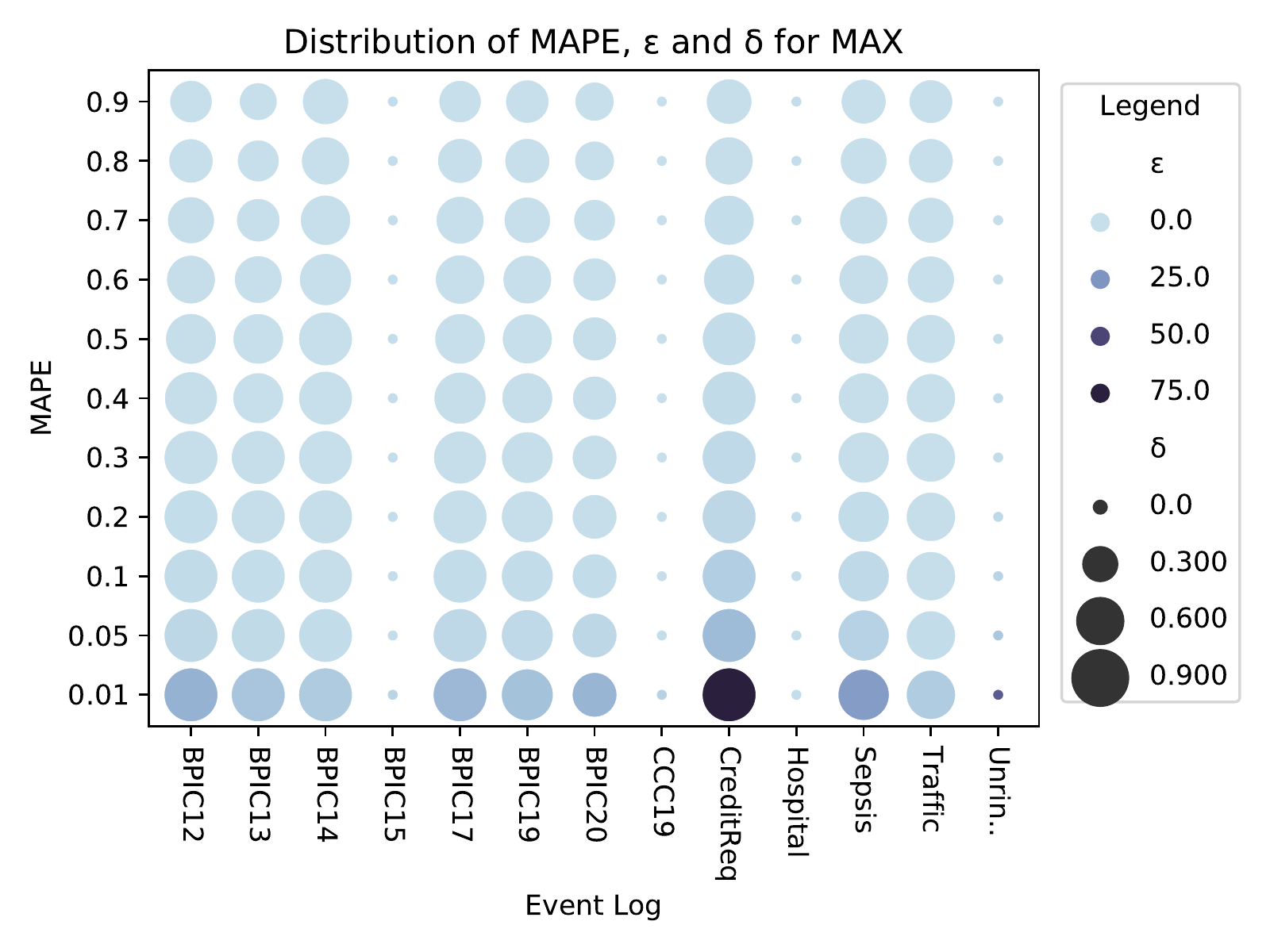}
		\caption{Max Time Difference Queries}
		\label{fig:alpha_max}
	\end{subfigure} 
	\bigskip
	\begin{subfigure}[b]{0.48\textwidth}
		\centering
		\includegraphics[width=.98\columnwidth]{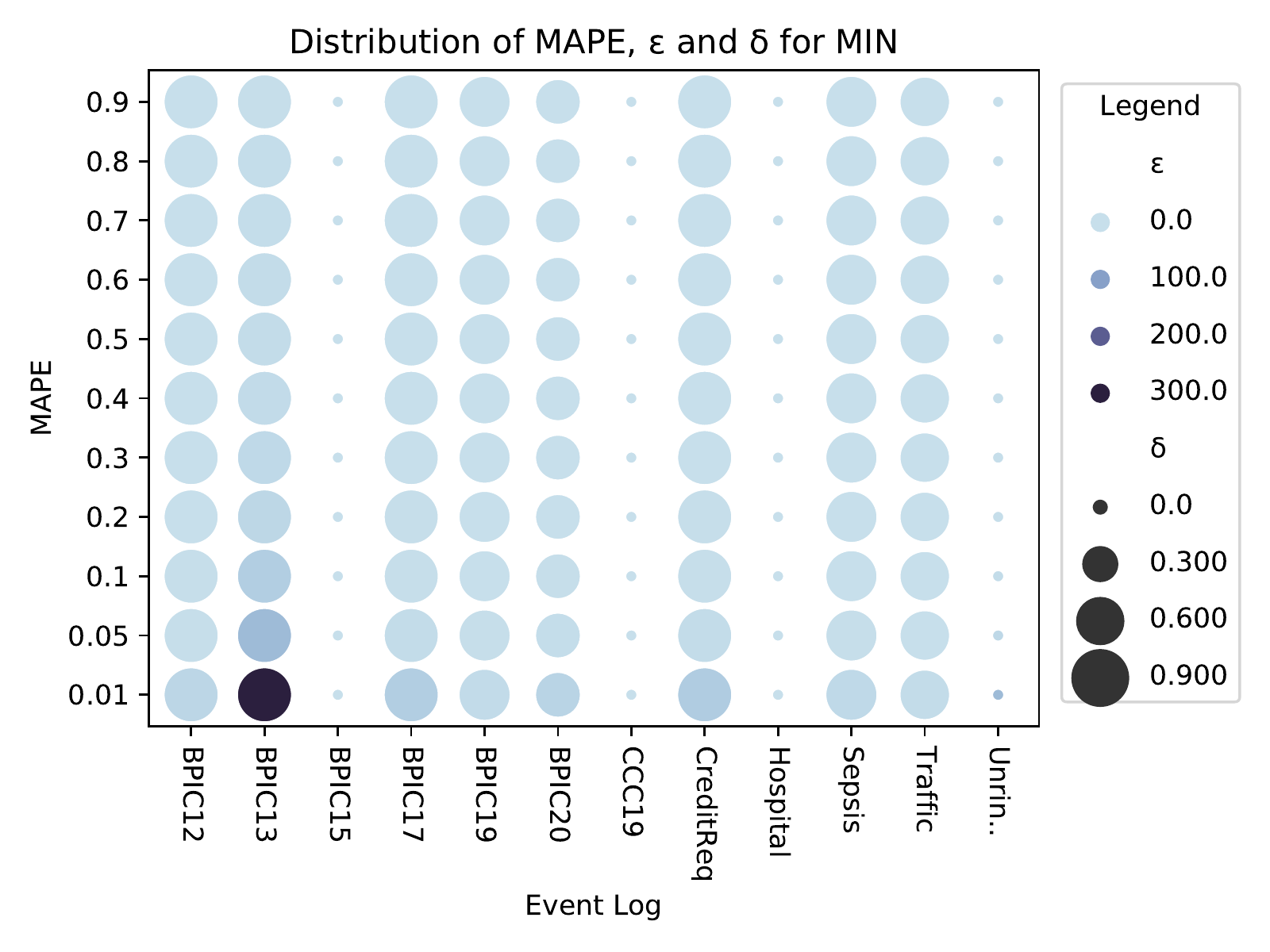}
		\caption{Min Time Difference Queries}
		\label{fig:alpha_min}
	\end{subfigure}\quad
	\begin{subfigure}[b]{0.48\textwidth}
		\centering
		\includegraphics[width=.98\columnwidth]{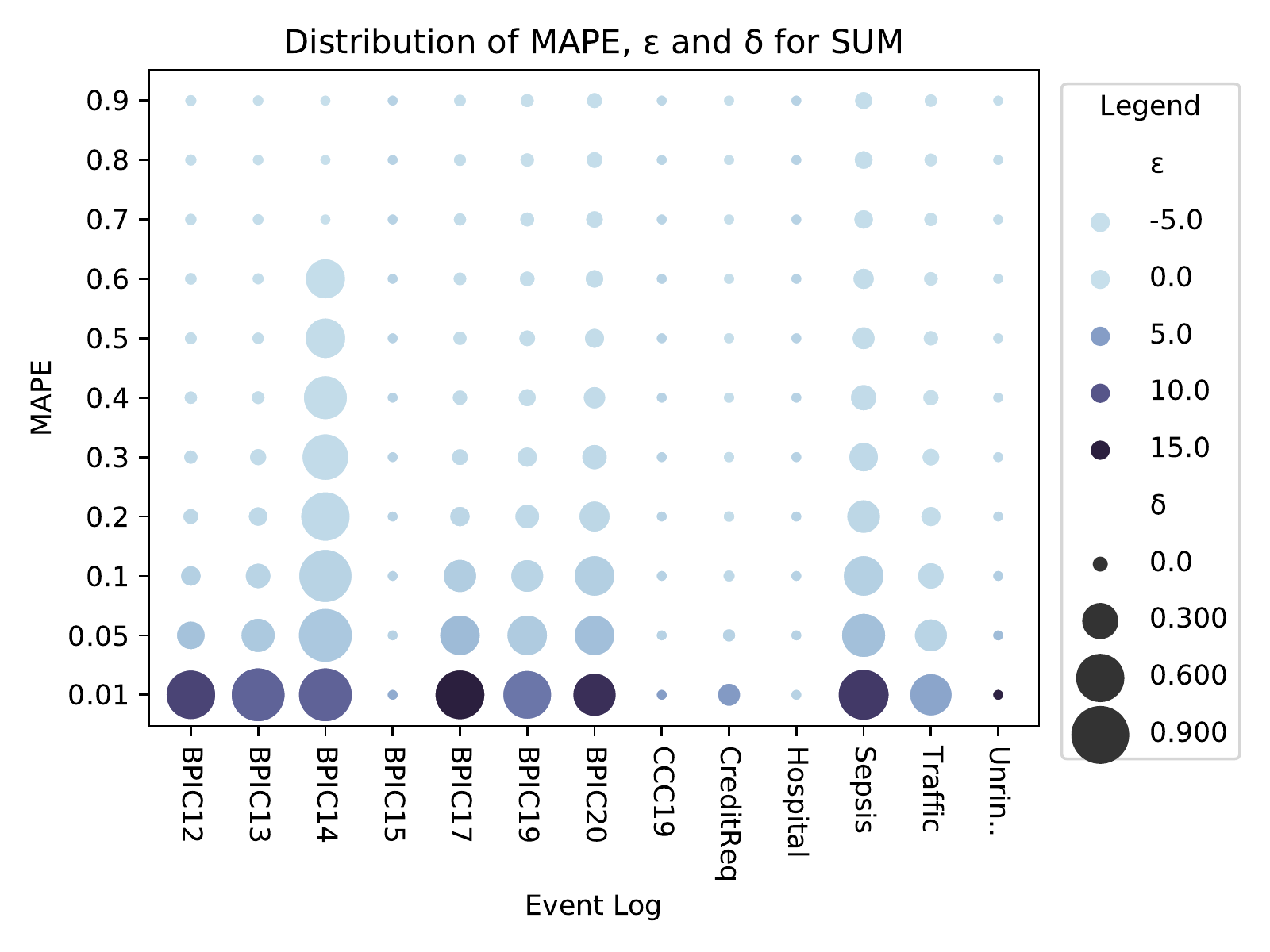}
		\caption{Total Time Difference Queries}
		\label{fig:alpha_sum}
	\end{subfigure}
	\bigskip
	\begin{subfigure}[b]{0.48\textwidth}
		\centering
		\includegraphics[width=.98\columnwidth]{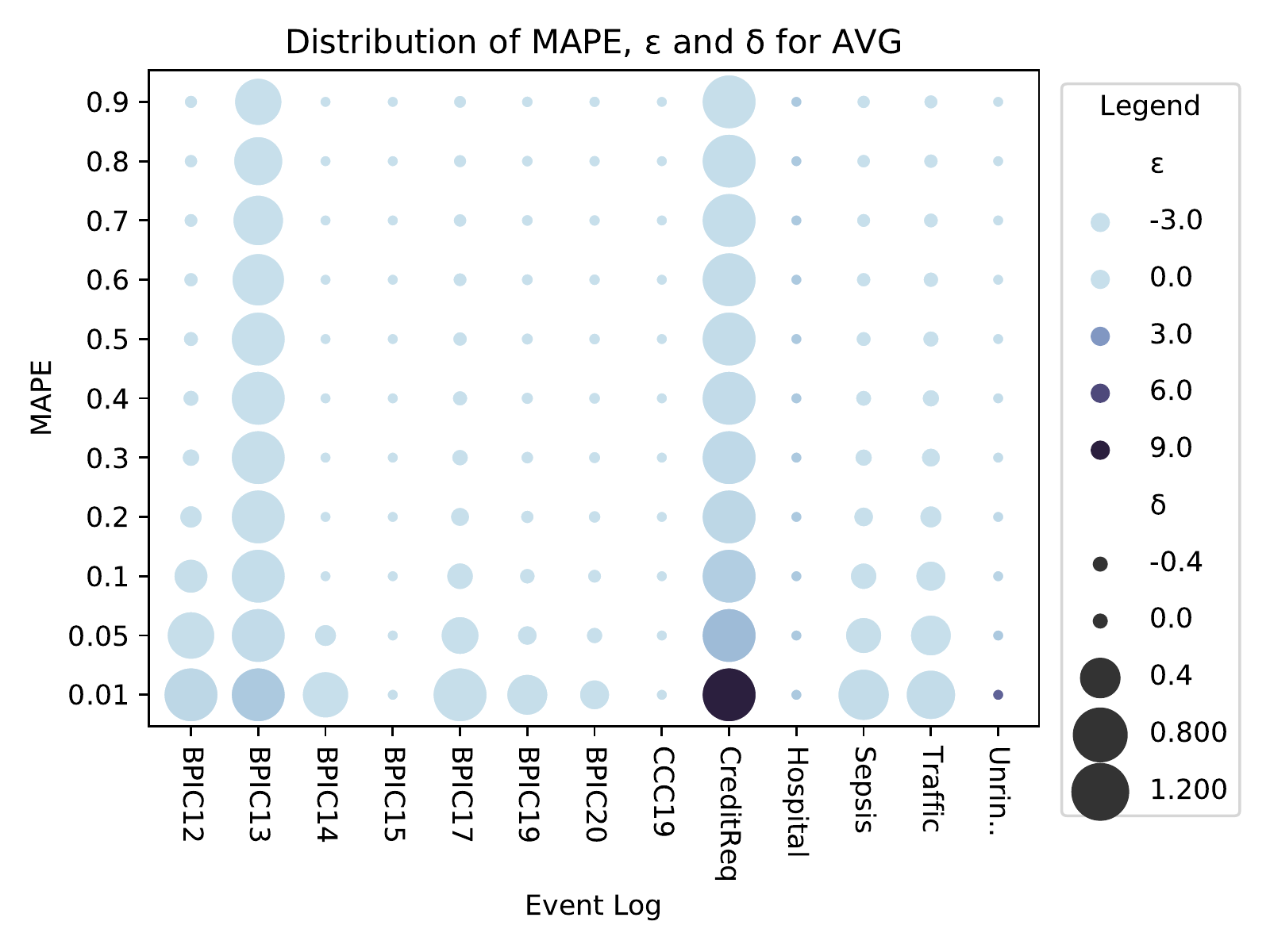}
		\caption{Average Time Difference Queries}
		\label{fig:alpha_avg}
	\end{subfigure}
	
	\caption{ MAPE's effect on $\epsilon$ and $\delta$. The columns are the event logs, the rows are amount of error MAPE, the size of the circles is the guessing advantage $\delta$, and The color hue is the  $\epsilon$.}
	\label{fig:bubble_alpha}
\end{figure*}

In problem~\ref{int:prob:2}, a process map publisher has a maximum acceptable percentage error that he wants to maintain. The publisher needs to calculate the corresponding differential privacy parameter, $\epsilon$, and also the amount of risk associated with that percentage error. We use the MAPE to calculate the $\alpha$ parameter per edge using Eq.~(\ref{eqn:alpha_per_edge}). Then, we use Eq.~(\ref{eqn:epsilon_laplace_per_edge}) to calculate $\epsilon$ for every edge separately.Given $\epsilon$, we estimate the guessing advantage probability that an attacker would gain after the disclosure, $\delta_{ij}$ of every edge that corresponds to the given error for both frequency and time queries. We calculate $\delta$ for frequency queries using Eq. (\ref{eq:delta_freq}), and for the time queries, we use Eq.~(\ref{eq:delta_time}) to calculate the $\delta$ associated with every instance of a directly-follows relation, then we take the maximum risk per edge using Eq.~(\ref{eq:delta_time_2_1}). We report the effect of utility loss level (the percentage error) on both $\epsilon$ and risk $\delta$.

 In Figure~\ref{fig:bubble_alpha}, we use the bubble heat maps, where the columns of every figure represent the event logs, the rows are different values for the amount of error MAPE. The size of the circles is the resulted risk $\delta$. The colour hue is the differential-privacy parameter, $\epsilon$, the darker the intensity, the lower the amount of injected noise. As we use different $\epsilon$ value across edges, we report the median $\epsilon$ across edges of event logs. The effect of input desired error appears through the columns. The more the permitted input error, the less the $\epsilon$ (more noise that would be added). The amount of noise injected depends on the aggregation and the input MAPE, e.g. in Figure~\ref{fig:alpha_sum}, the values of $\epsilon$ are large at the bottom.

\begin{figure*}[t!]
	\centering
		\begin{subfigure}[b]{0.48\textwidth}
		\centering
		\includegraphics[width=.98\columnwidth]{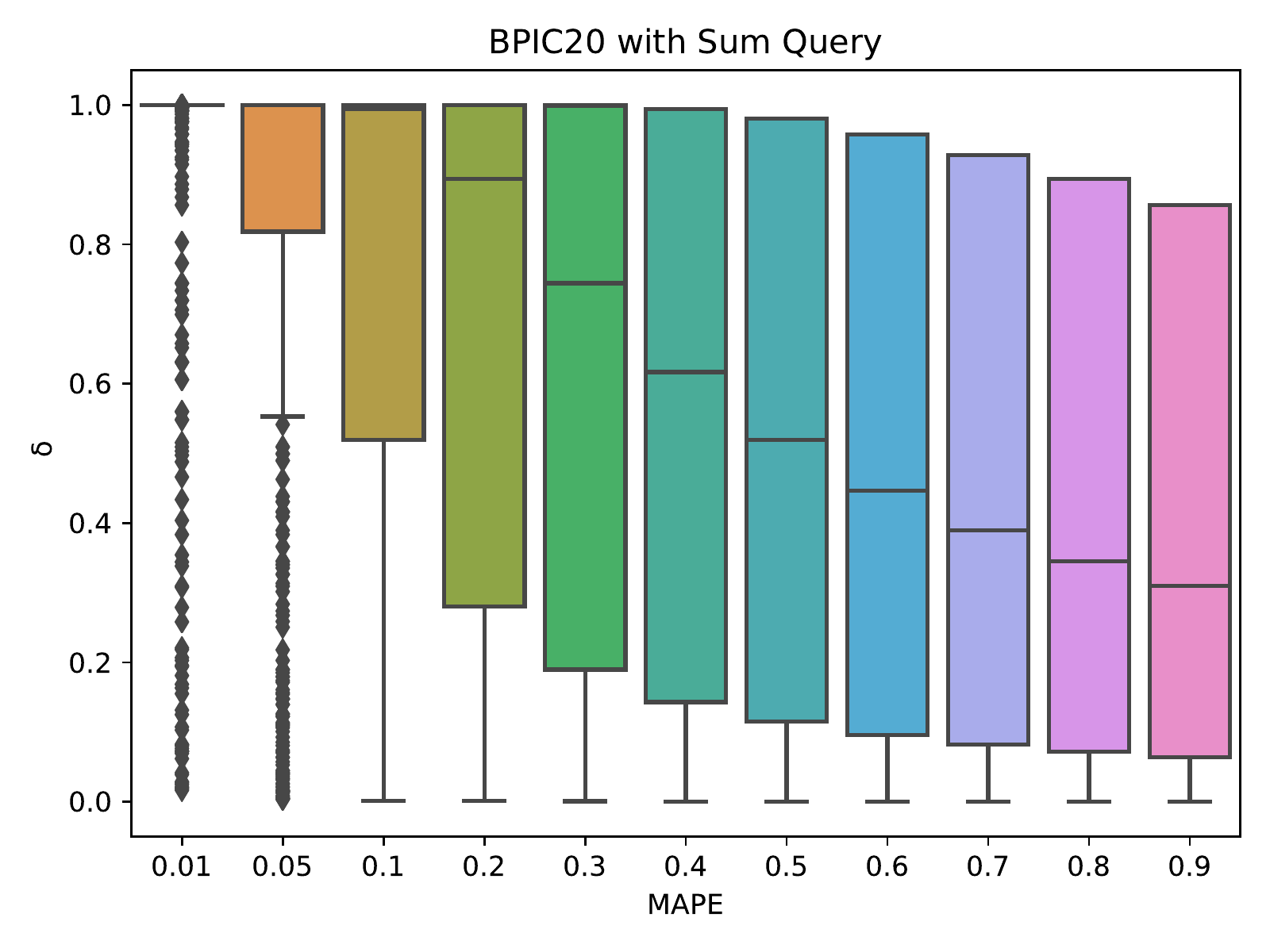}
		\caption{$\delta$ Distribution across the edges of BPIC20 with Sum Query}
		\label{fig:bpic20_distribution}
	\end{subfigure}\quad 
	\begin{subfigure}[b]{0.48\textwidth}
		\centering
		\includegraphics[width=.98\columnwidth]{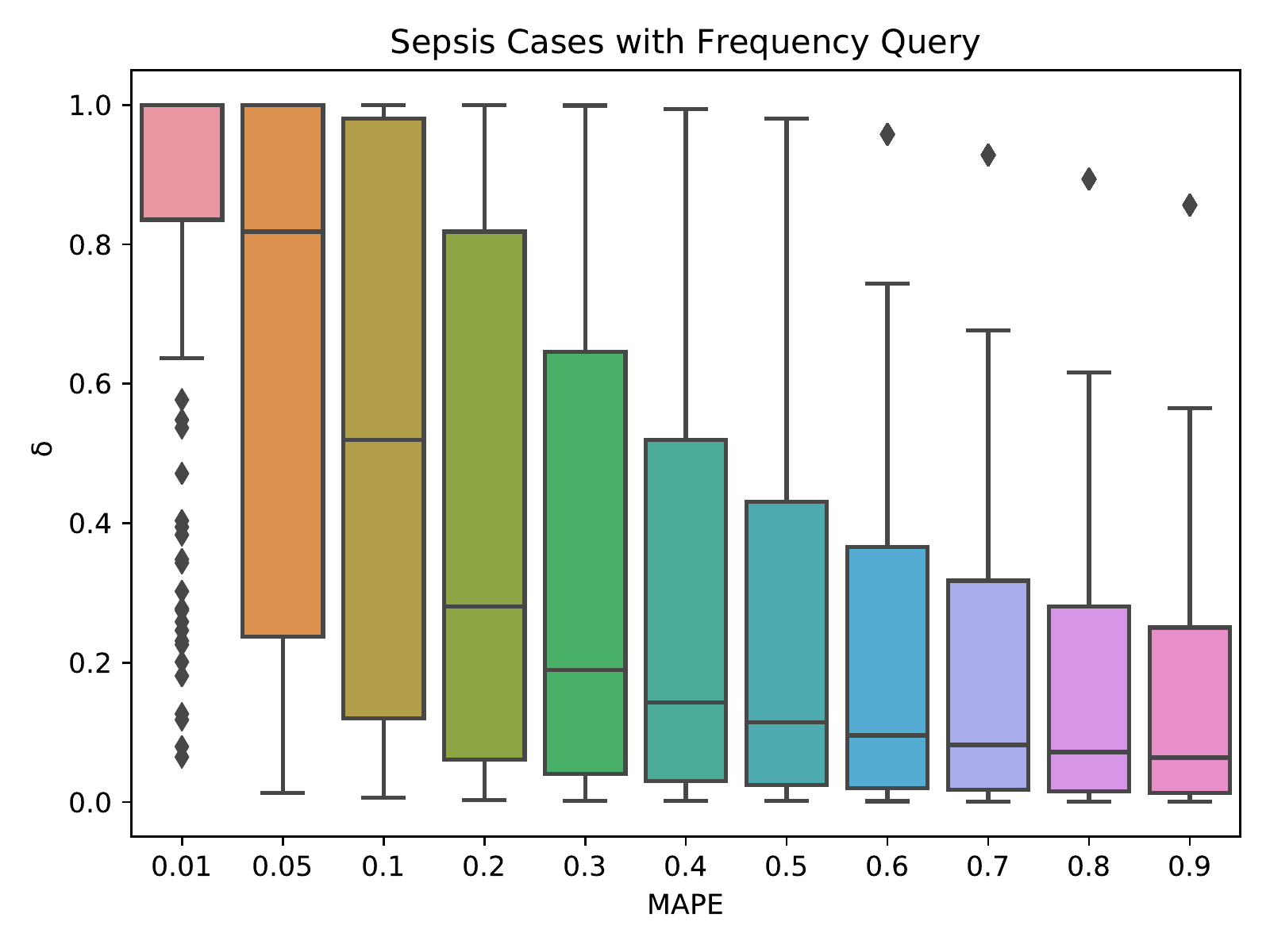}
		\caption{$\delta$ Distribution across the edges of Sepsis Cases with Frequency Query}
		\label{fig:sepsis_distribution}
	\end{subfigure}
	\caption{$\delta$ Distribution across the edges of Event Logs (Log Scale)}
	\label{fig:alpha_distribution}
\end{figure*}

The relation between the permitted percentage error, MAPE, and the guessing advantage, $\delta$, is shown in Figure~\ref{fig:bubble_alpha}. The median guessing advantage across edges, size of the circles, decreases with increasing the input MAPE. We can conclude that the relation between utility and risk depends on the data distribution and the aggregation, as explained in Section~\ref{sec:balance}. In most of the cases, the model maintains high utility levels with low risk. The risk differs across edges. The risk distribution across edges is shown in Figures~\ref{fig:bpic20_distribution} and ~\ref{fig:sepsis_distribution} for both BPIC20 and Sepsis Cases event logs, as a result of sum and frequency aggregates, respectively. It is observable in some event logs, due to very few risky edges, the risk attached with publishing the DFG is high as we consider the maximum risk. For instance, in Figure~\ref{fig:sepsis_distribution} the median $\delta$ for the Sepsis cases is 0.3 for a MAPE of 0.2, but the maximum risk across edges equals to 1, which happens in very few edges. The distribution of the risk over edges of all event logs is available in the supplementary material~\cite{supplementaryMaterial}. 
Combining the proposed model with a mechanism to suppress the risky events and edges would enhance the balance between risk and utility.

\subsection{Execution Time Experiment}

\begin{figure}[t!]
	\centering
		\includegraphics[width=.7\columnwidth]{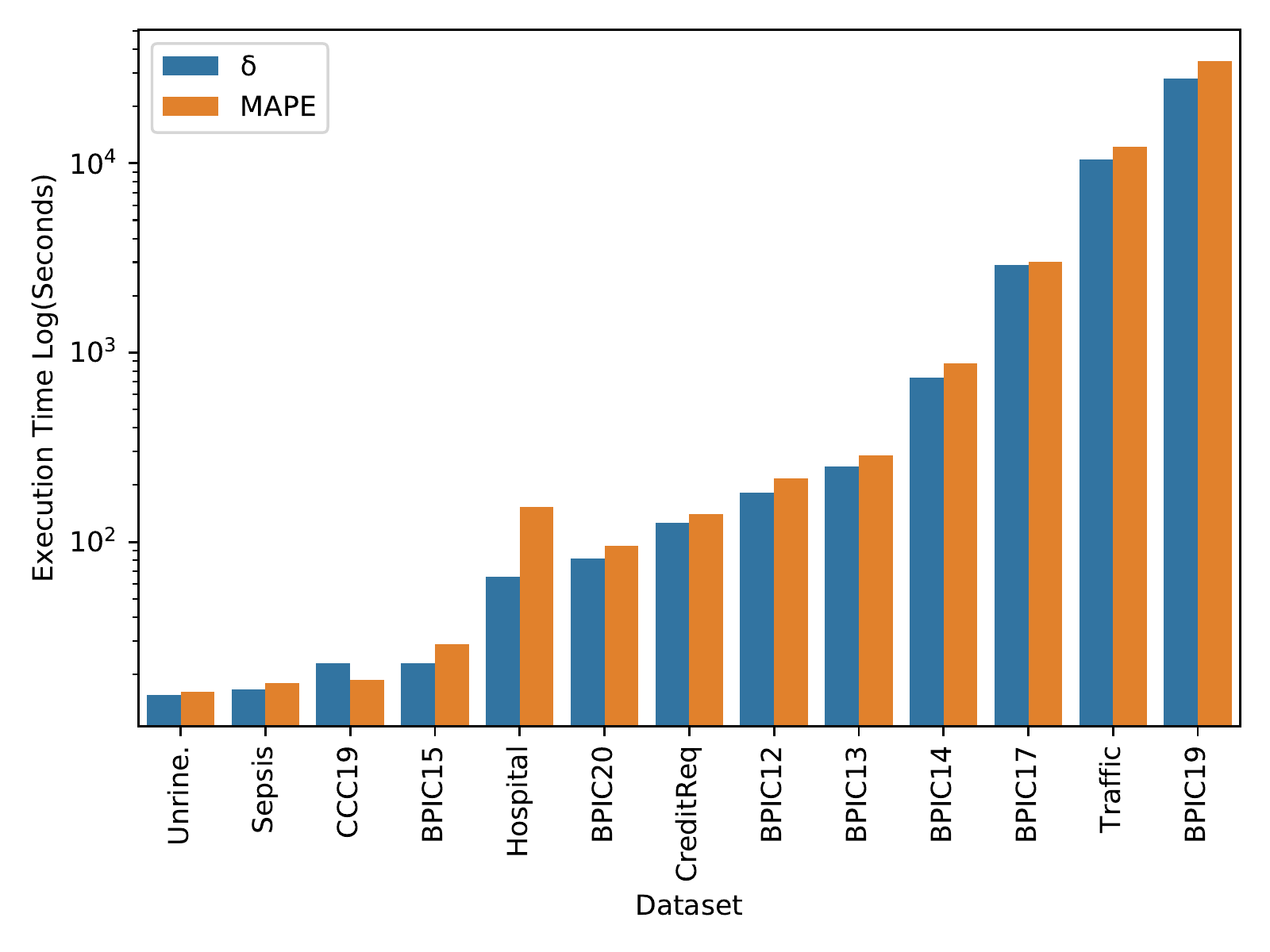}
	\caption{Execution Time Experiment}
	\label{fig:execution_time}
\end{figure}

We conducted an execution time experiment to measure the wall-to-wall execution time of the proposed method. We perform the experiment using 4 cores CPU. In Figure~\ref{fig:execution_time}, we show the execution time experiment of one the average aggregate type across the event logs, and we use the log scale as the difference between execution time experiments of event logs is large. From Figure~\ref{fig:execution_time}, we can see that the execution time has a different behavior based on the occurrence frequency of a directly follows relation, i.e., the number of edges and the number of occurrences of the relation. For instance, the Credit Requirement event log has a smaller number of edges than Sepsis Cases but more occurrences of edges, which causes the execution time of Credit Requirement to be longer than the execution time of Sepsis Cases. On the other hand, Credit Requirement contains fewer edges than Road Traffic event log, but the Road Traffic has a significantly larger number of events per edge so that it has a longer execution time. The edge calculations are heavily parallelizable, e.g., we run a single experiment for the BPIC 2014 with a 10 cores CPU, the execution time dropped to 5 minutes and 39 seconds, from 14 minutes and 34 seconds with a 4 cores CPU.

\subsection{Threats to Validity}

In the proposed method, we consider only the noise drawn from a Laplacian distribution, and that is for simplifying the calculations of the model. Other noise distributions may produce better results with some event logs, depending on the attributes of the event logs and the values on the edges. Some query operations might have a high-prior knowledge measure that the method cannot handle, especially with the time difference aggregate queries. Furthermore, in the absence of distribution of input values over the edge, e.g., frequency annotated DFGs, we considered the worst-case prior probability, which may result in higher noise, as explained in Section~\ref{sec:balance}.
We have performed our experiments on only publicly available event logs. However, to mitigate this risk, we selected 13 event logs using carefully justified criteria, in such a way that the selected logs are representative of a broader pool of logs.


\section{Conclusion and Future Work}
\label{sec:conclusion}
In this article, we proposed a method to protect the disclosure of DFGs extracted from business process event logs in order to strike a balance between disclosure risk and utility loss under a differential privacy model.  
Specifically, the proposed method estimates an optimal value of the $\epsilon$ parameter used by differentially private mechanisms, given a maximum level of disclosure risk -- captured via an adversary guessing advantage measure -- or a maximum level of  utility loss -- captured via absolute percentage error.
The method addresses two use cases. 
In the first one, the user specifies a maximum tolerable guessing advantage and the method estimates an $\epsilon$ value that fulfills the specified guessing advantage, while minimizing the percentage error. 
In the second use case, the user specifies a maximum tolerable percentage error and the method estimates an $\epsilon$ value that fulfills the specified percentage error, while minimizing the guessing advantage that the disclosed DFG provides to an adversary. 


The method proposed in this article works entirely by adding differentially private noise to the weights of the arcs of the DFG. In other words, the method alters the frequency of each arc or the aggregate time difference attached to each arc, depending on the type of DFG being disclosed. 
In some scenarios, particularly when a given directly-follows relation is infrequent or in the presence of outliers in the time differences, it may be more effective to suppress a given arc in the DFG altogether. 
For example, in Figure~\ref{fig:sepsis_distribution}, we notice that a handful of edges have high risk due to presence of outliers. If we simply removed the 10\% most frequent edges, the maximum disclosure risk measure would drop below 8\%. Accordingly, a direction for future work is to extend the proposed method in order to support two types of noise-injection operations: altering the weights in the arcs of the DFG (as in the current method) and suppressing arcs of the DFG. This extended method is likely to provide higher flexibility when it comes to achieving trade-offs between utility loss and disclosure risk.

In this article, we focused on the problem of disclosing the DFG of an event log. While DFGs are arguably one of the most common types of abstractions used in process mining, other abstractions are also commonly used, including the distribution of distinct activity traces of an event log, as highlighted in~\cite{mannhardt2019privacy}. Another avenue for future work is to design methods for protecting the disclosure of other types of log abstractions besides DFGs.


\medskip\noindent\textbf{Acknowledgments.} This research is funded by European Social Funds (ESF) via the IT Academy Programme.

 \bibliographystyle{splncs04}
 \bibliography{Amun}

\end{document}